\documentclass[acmsmall]{acmart}

\AtBeginDocument{%
  }

\setcopyright{cc}
\setcctype{by}
\acmDOI{10.1145/3720431}
\acmYear{2025}
\acmJournal{PACMPL}
\acmVolume{9}
\acmNumber{OOPSLA1}
\acmArticle{97}
\acmMonth{4}
\received{2024-10-16}
\received[accepted]{2025-02-18}

\usepackage[normalem]{ulem}

\usepackage{xspace}
\usepackage{enumitem}
\usepackage{soul}
\usepackage{xcolor}
\usepackage{comment}
\usepackage{textcomp}

\usepackage{url}

\usepackage{float}
\usepackage{graphicx}

\usepackage{xspace}
\usepackage{listings}
\usepackage{placeins}
\usepackage{makecell}

\usepackage{tikz}
\usetikzlibrary{positioning}
\usepackage{multirow}
\usepackage{setspace}
\usepackage{subcaption}
\usepackage{tablefootnote}
\usepackage{pifont}

\usepackage{cleveref}
\crefformat{section}{#2\S#1#3}
\Crefformat{section}{#2\S#1#3}
\crefformat{subsection}{#2\S#1#3}
\Crefformat{subsection}{#2\S#1#3}
\crefname{line}{Line}{Lines}
\Crefname{line}{Line}{Lines}
\crefformat{line}{#2Line~#1#3}
\Crefformat{line}{#2Line~#1#3}
\crefname{listing}{Listing}{Listings}
\Crefname{listing}{Listing}{Listings}
\crefname{lstlisting}{Listing}{Listings}
\Crefname{lstlisting}{Listing}{Listings}
\crefformat{listing}{#2Listing~#1#3}
\Crefformat{listing}{#2Listing~#1#3}
\crefname{figure}{Figure}{Figures}
\Crefname{figure}{Figure}{Figures}
\crefformat{figure}{#2Fig.~#1#3}
\Crefformat{figure}{#2Fig.~#1#3}

\crefrangeformat{section}{#3\S#1#4--#5#2#6}
\Crefrangeformat{section}{#3\S#1#4--#5#2#6}
\crefrangeformat{subsection}{#3\S#1#4--#5#2#6}
\Crefrangeformat{subsection}{#3\S#1#4--#5#2#6}

\crefname{table}{Table}{Tables}
\Crefname{table}{Table}{Tables}

\newcommand{\refline}[1]{line~\ref{#1}}

\usepackage{listings}
\usepackage{xcolor}
\usepackage{soul}

\definecolor{yelloworange}{RGB}{255, 200, 0}
\sethlcolor{yelloworange}

\definecolor{darkgreen}{RGB}{0, 128, 0}
\definecolor{commentgreen}{rgb}{0, 0.5, 0}

\lstdefinelanguage
[x64]{Assembler}     %
[x86masm]{Assembler} %
{morekeywords={CDQE,CQO,CMPSQ,CMPXCHG16B,JRCXZ,LODSQ,MOVSXD, %
		POPFQ,PUSHFQ,SCASQ,STOSQ,IRETQ,RDTSCP,SWAPGS, %
		rax,rdx,rcx,rbx,rsi,rdi,rsp,rbp, %
		r8,r8d,r8w,r8b,r9,r9d,r9w,r9b}} %

\lstdefinestyle{customc}
{
	belowcaptionskip=-0.5\baselineskip,
	breaklines=true,
	captionpos=b,                    %
	language=C,
	showstringspaces=false,
	basicstyle=\fontsize{8}{7}\selectfont\bfseries\ttfamily,
	keywordstyle=\color{black},
	commentstyle=\itshape\color{gray!70!black},
	identifierstyle=\color{black},
	stringstyle=\color{red!70!black},
	emph={static,volatile,double,float,signed,unsigned,int,void,size_t,char, key_t, value_t,STORE, FLUSH,FENCE,uint64_t,struct},
	emphstyle={\color{olive}},
	numbersep=8pt,
}

\lstdefinestyle{customnew}
{
    backgroundcolor=\color{lightgray!10}, %
    commentstyle=\color{darkgreen}\textit, %
    keywordstyle=\color{blue}\bfseries, %
    numberstyle=\tiny\color{gray}, %
    stringstyle=\color{orange}, %
    basicstyle=\ttfamily\footnotesize, %
    breakatwhitespace=false, %
    breaklines=true, %
    captionpos=b, %
    keepspaces=true, %
    numbers=left, %
    numbersep=5pt, %
    showspaces=false, %
    showstringspaces=false, %
    showtabs=false, %
    tabsize=2, %
    frame=single, %
    rulecolor=\color{black}, %
    aboveskip=1.5em, %
    belowskip=1.5em, %
    frameround=tttt, %
}

\usepackage[
  vlined,
  linesnumbered,
]{algorithm2e}

\SetKwComment{Comment}{$\triangleright$\ }{}

\SetCommentSty{mycommfont}
\SetAlgoLined

\newif\ifdraft
 \draftfalse

\newcommand{\sys}{\textsc{Path}\-\textsc{finder}\xspace}

\newcommand{\stt}[1]{\texttt{\small #1}\xspace}

\newcommand{\add}[1]{\ifdraft{\color{blue}#1}\else{#1}\fi}

\renewcommand{\paragraph}[1]{\textbf{\itshape #1.}}

\newcommand{\totalServerCCBugs}{18\xspace}
\newcommand{\newServerCCBugs}{7\xspace}
\newcommand{\confirmedServerBugs}{2\xspace}
\newcommand{\numServerApps}{8\xspace}

\newcommand{\totalMMIOServerCCBugs}{10\xspace}
\newcommand{\newMMIOServerCCBugs}{5\xspace}
\newcommand{\totalMMIOMicroCCBugs}{102\xspace}
\newcommand{\newMMIOMicroCCBugs}{49\xspace}

\newcommand{\numPOSIXApps}{3\xspace}

\newcommand{\numMMIOServerApps}{5\xspace}
\newcommand{\numMMIOMicroApps}{16\xspace}

\newcommand{\totalPOSIXCCBugs}{\add{8}\xspace}
\newcommand{\newPOSIXCCBugs}{2\xspace}

\newcommand{\POSIXRatio}{4$\times$\xspace}

\newcommand{\MMIORatio}{\JaaruRatio}
\newcommand{\JaaruRatio}{8$\times$\xspace}

\newcommand{\POSIXTime}{2 hours\xspace}
\newcommand{\MMIOTime}{2 hours\xspace}

\newcommand{\newRecipeBugs}{17\xspace}
    
\newcommand{\newPMDKBugs}{6\xspace}
\newcommand{\allServerBugs}{6\xspace}
    \newcommand{\newMCDBugs}{2\xspace}
    \newcommand{\newRedisBugs}{1\xspace}

\newcommand{\newKVBugs}{26\xspace}

\newcommand{\numPmdkMicroApps}{6\xspace}
\newcommand{\numWitcherApps}{5\xspace}
\newcommand{\numRecipeApps}{5\xspace}

\newcommand{\posixBased}{POSIX-based\xspace}
\newcommand{\mmioBased}{MMIO-based\xspace}
\newcommand{\squintPOSIX}{\sys{}\xspace}
\newcommand{\squintMMIO}{\sys{}\xspace}

\newcommand{\algrule}[1][.7pt]{\par\vskip.5\baselineskip\hrule height #1\par\vskip.5\baselineskip}

\begin{document}

\title{Scalable and Accurate Application-Level Crash-Consistency Testing via Representative Testing}

\author{Yile Gu}
\authornote{Both authors contributed equally to this research.}
\orcid{0009-0009-8292-7232}
\affiliation{%
  \institution{University of Washington}
  \city{Seattle}
  \country{USA}
}
\email{yilegu@cs.washington.edu}

\author{Ian Neal}
\authornotemark[1]
\orcid{0000-0001-9721-781X}
\affiliation{%
  \institution{University of Michigan}
  \city{Ann Arbor}
  \country{USA}
}
\affiliation{%
  \institution{Veridise}
  \city{Austin}
  \country{USA}
}
\email{iangneal@umich.edu}

\author{Jiexiao Xu}
\orcid{0009-0004-2752-5664}
\affiliation{%
  \institution{University of Washington}
  \city{Seattle}
  \country{USA}
}
\email{jiexiao@cs.washington.edu}

\author{Shaun Christopher Lee}
\orcid{0009-0000-4521-6778}
\affiliation{%
  \institution{University of Washington}
  \city{Seattle}
  \country{USA}
}
\email{shauncl8@cs.washington.edu}

\author{Ayman Said}
\orcid{0009-0002-5409-4367}
\affiliation{%
  \institution{University of Michigan}
  \city{Ann Arbor}
  \country{USA}
}
\email{sayman@umich.edu}

\author{Musa Haydar}
\orcid{0009-0008-5137-587X}
\affiliation{%
  \institution{University of Michigan}
  \city{Ann Arbor}
  \country{USA}
}
\email{musah@umich.edu}

\author{Jacob Van Geffen}
\orcid{0009-0007-7468-4205}
\affiliation{%
  \institution{Veridise}
  \city{Austin}
  \country{USA}
}
\email{jacob@formalstack.com}

\author{Rohan Kadekodi}
\orcid{0000-0002-1213-0342}
\affiliation{%
  \institution{University of Washington}
  \city{Seattle}
  \country{USA}
}
\email{rohankad@cs.washington.edu}

\author{Andrew Quinn}
\orcid{0000-0002-0785-4119}
\affiliation{%
  \institution{University of California at Santa Cruz}
  \city{Santa Cruz}
  \country{USA}
}
\email{aquinn1@ucsc.edu}

\author{Baris Kasikci}
\orcid{0000-0001-6122-8998}
\affiliation{%
  \institution{University of Washington}
  \city{Seattle}
  \country{USA}
}
\email{baris@cs.washington.edu}

\renewcommand{\shortauthors}{Y. Gu, I. Neal, J. Xu, S. C. Lee, A. Said, M. Haydar, J. V. Geffen, R. Kadekodi, A. Quinn, and B. Kasikci}

\begin{abstract}

Crash consistency is essential for applications that must persist data.
Crash-consistency testing has been commonly applied to find crash-consistency bugs in applications.
The crash-state space grows exponentially as the number of operations in the program increases, necessitating techniques for pruning the search space.
However, state-of-the-art crash-state space pruning is far from ideal.
Some techniques look for known buggy patterns or bound the exploration for efficiency, but they sacrifice coverage and may miss bugs lodged deep within applications.
Other techniques eliminate redundancy in the search space by skipping identical crash states, but they still fail to scale to larger applications.

In this work, we propose \emph{representative testing}: a new crash-state space reduction strategy that achieves high scalability and high coverage.
Our key observation is that the consistency of crash states is often correlated, even if those crash states are not identical.
We build \sys, a crash-consistency testing tool that implements an \emph{update behaviors-based} heuristic to approximate a small set of representative crash states.

We evaluate \sys on \posixBased and \mmioBased applications, where it finds \totalServerCCBugs (\newServerCCBugs new) bugs across \numServerApps production-ready systems.
\sys scales more effectively to large applications than prior works and finds \POSIXRatio more bugs in \posixBased applications and \MMIORatio more bugs in \mmioBased applications compared to state-of-the-art systems.

\end{abstract}



\maketitle

\section{Introduction}
\label{sec:introduction}
\label{sec:intro}

Crash consistency refers to an application's ability to maintain a consistent state for its data in the presence of an unexpected system crash.
For systems that require data persistence, such as storage engines~\cite{leveldb, rocksdb,wiredtiger}, cloud controllers~\cite{hadoop,zookeeper}, and HPC libraries~\cite{hdf5,NetCDF}, failure to ensure crash consistency could lead to detrimental consequences such as data corruption and system errors.

Crafting crash-consistent applications is challenging.
An application typically does not directly interact with storage devices, but instead runs on top of file systems that provide the POSIX interface~\cite{posix} and issues syscalls or uses memory-mapped IO (MMIO) to persist data.
For performance benefits, file systems may buffer updates in-memory, requiring application developers to issue explicit ordering calls like \emph{fsync} to flush the data to the disk~\cite{ALICE}. 

Such explicit ordering calls make reasoning about the persistent state of an application easier, as the number of possible crash states is reduced.
However, these ordering calls are often expensive, as they limit the ability of the underlying storage system to buffer and batch updates.
Therefore, as developers strive to design efficient systems, they can quickly become difficult to reason about, leading to crash-consistency bugs that are challenging for developers to find without assistance.

Software testing has been commonly used to find crash-consistency bugs in applications.
Crash-consistency testing tools find bugs by first generating all of an execution's possible crash states and exhaustively applying (or not applying) persistent updates that are in an intermediate state (i.e., those that have not been explicitly synced).
The tools then test the application with these crash states to see if recovering from them yields an inconsistent state.
For example, file systems can flush the data for write syscalls that are issued to different files in any order~\cite{Ferrite}. So, the set of possible crash states for two such writes includes four states: when neither file is updated, both are updated, or one or the other file is updated.
Real-world applications generate an extremely large number of crash states during a single program execution, making application-level crash-consistency testing of this nature challenging.
In the worst case, the crash-state space could grow exponentially as the number of operations in the program increases.
This problem, known as "crash-state space explosion"~\cite{fu2021witcher,PACE}, prevents crash-consistency testing tools from testing all possible crash states.

Existing crash-consistency testing techniques manage the crash-state space explosion problem by testing a subset of crash states that can be produced by an application~\cite{FiSC, leesatapornwongsa2014samc,gorjiara2021jaaru, COFI, PACE, torturingdb, CrashMonkey, fu2021witcher, liu2019pmtest,engler2001bugs,neal2020agamotto, jiang2016crash}.  
However, such techniques either sacrifice coverage, i.e., they do not find crash states that lead to inconsistency, or efficiency, i.e., they test too many crash states and cannot scale to large programs.   
For example, many techniques prune the crash-state space by looking for known likely buggy patterns~\cite{PACE, torturingdb, CrashMonkey, fu2021witcher, liu2019pmtest, engler2001bugs,neal2020agamotto} or by bounding exploration to an arbitrary time or depth limit~\cite{COFI, jiang2016crash}.
These tools are efficient, but cannot find bugs that do not match known patterns or bugs that occur deep within a program's execution.
Other techniques~\cite{FiSC, leesatapornwongsa2014samc,gorjiara2021jaaru} retain coverage and attempt to improve efficiency by eliminating redundancy in the crash-state space.  
However, these solutions only identify redundancy for crash states that are \emph{identical}, which greatly reduces their scalability.  
For example, while Jaaru~\cite{gorjiara2021jaaru} uses dynamic partial order reduction (DPOR)~\cite{flanagan2005dynamic} to scale persistent memory crash-consistency testing better than prior exhaustive approaches~\cite{lantz2014yat,pmreorder}, it can still only scale to 1,000-line libraries.

Our key insight is that the consistency of crash states is often \emph{correlated} (i.e. they expose crash-consistency bugs with the same root cause), even when those crash states are not identical.
This insight stems from our observation that the root causes of many crash-consistency bugs are not related to the data values written by the operations in a program, but to the unexpected ordering of operations persisted to disk that produces an inconsistent crash state. 
In other words, crash-consistency bugs arise when old and new data are incorrectly combined, regardless of the precise values contained in the old and new data.
Based on this observation, we define the notion of a \emph{representative set of crash states}: \textbf{the consistency of a set of crash states, $\mathbf{C_1}$, is implicitly tested by testing the consistency of another \emph{representative} set of crash states, $\mathbf{C_2}$, if all of the crash states in $\mathbf{C_1}$ are correlated with the crash states in $\mathbf{C_2}$.}

Based on our insight, we propose a new state-space reduction technique, \emph{representative testing}.
Representative testing prunes the crash-state space by identifying and testing a small number of representative sets of crash states.
If a set of representative crash states is found to not contain any crash-consistency bugs, the \emph{represented} sets of crash states are safe to prune, as we can conclude they also do not contain bugs.

Perfectly identifying all possible representative crash states and all the crash states they represent would require testing all crash states from a given execution, but such an approach would be computationally infeasible.
Therefore, we propose and implement an \emph{update behaviors-based} heuristic to approximate correlated crash states without performing exhaustive crash-consistency testing.
An update behavior is a sequence of operations that are semantically related (i.e. they are performing a task that is meaningful from application's point of view).
The exact definition of the task is dependent on applications. 
For example, a database application could perform the task of background compaction, or a persistent BTree could perform the task of removing one leaf node.
Our heuristic identifies and leverages update behaviors to group correlated crash states. Intuitively, behaviors in an application that aim to perform the same task generally must enforce the same consistency conditions. Thus, identifying bugs in one instance of an update behavior is sufficient to identify bugs in any behavior that is attempting to perform the same task.

We perform two case studies on how update behaviors-based heuristic can be implemented: one for \posixBased applications and one for \mmioBased applications.
An update behavior is defined at the granularity of how applications persist data to durable storage.
\posixBased applications persist updates to disk using syscalls, while \mmioBased applications directly persist updates via memory stores to memory-mapped files.
Therefore, each operation in an update behavior is a syscall in \posixBased applications and a memory operation in \mmioBased applications.
For \posixBased applications, we observe that developers commonly group semantically related syscalls together under the same function.
As a result, we split a trace of syscalls issued by the \posixBased program into update behaviors based on their function (see \Cref{subsec:posix-case-study} for a detailed case study).
Similarly, for \mmioBased applications, we find that semantically-related data fields are commonly grouped together under the same data type.
We split a trace of memory updates collected from the \mmioBased program into update behaviors based on the affected data type (see \cref{subsec:mmio-case-study} for a detailed case study). 

We present \sys{}, a scalable and accurate application-level crash-consistency testing tool using representative testing.  
\sys{} implements two versions of our update behaviors-based heuristic mentioned above for \posixBased and \mmioBased application crash-consistency testing. 
Inspired from prior work on file system dependencies~\cite{Featherstitch}, \sys{} provides a universal abstraction called a \emph{Persistence Graph} to express the ordering constraints that are enforced by the application, where the nodes are program updates (i.e. syscalls in \posixBased applications, stores in \mmioBased applications) and edges are "happens-before" relations (enforced by \stt{fsync} and \stt{fdatasync} in \posixBased
 applications, and by memory fence and \stt{msync} in \mmioBased applications).
To test an application, \sys{} first traces an execution of the application to identify all program updates. 
It then uses the update behavior heuristic mentioned above to identify the set of representative behaviors.
\sys then checks the crash states resulting from these behaviors, using DPOR to prune redundant states.

\squintPOSIX is able to find \totalPOSIXCCBugs bugs across \numPOSIXApps widely-tested \posixBased large database applications, where \newPOSIXCCBugs are new bugs.
In addition, \squintMMIO finds \totalMMIOServerCCBugs bugs across \numMMIOServerApps production-ready \mmioBased applications with \newMMIOServerCCBugs new bugs.
In terms of scalability, we compare \squintPOSIX against ALICE~\cite{ALICE}, a widely used crash-consistency testing tool for \posixBased applications.
On average, \squintPOSIX finds \POSIXRatio more bugs under a time limit of \POSIXTime.
We also compare \squintMMIO against Jaaru~\cite{gorjiara2021jaaru}, the state-of-the-art DPOR-based model checking tool for \numMMIOMicroApps \mmioBased microbenchmarks.
\squintMMIO finds \MMIORatio more bugs than Jaaru under \MMIOTime time limit.
We believe our evaluation shows that representative testing is an effective method for scalable and accurate crash-consistency testing.

We summarize our contributions as follows:
\begin{itemize}
    \item We propose \emph{representative testing}, a state-space reduction method that reduces the crash-state testing space by eliminating the testing of crash-states that are likely to evince the same crash-consistency bugs.
    \item \add{We design and implement \sys{}\footnote{\sys{} is publicly available at \url{https://github.com/efeslab/Pathfinder}.}, a crash-consistency testing tool that applies representative testing to achieve high scalability and accuracy.}
    \item We demonstrate how the \emph{update behaviors-based} heuristic can be implemented for scalable and accurate crash-consistency testing of \posixBased and \mmioBased applications.
    \item We evaluate \sys{} and find that it finds \POSIXRatio more bugs in \posixBased applications and \MMIORatio more bugs in \mmioBased applications compared to baseline systems.
\end{itemize}

\section{Representative Testing}
\label{sec:rep-testing}

We describe representative testing in detail and how it can be applied to crash-consistency testing for \posixBased and \mmioBased applications.

\subsection{General Methodology}
\label{subsec:general-methodology}

Representative testing is a crash-consistency testing method that selectively tests representative program behaviors with an aim to provide \emph{efficient} and \emph{accurate} testing.
Representative testing is based on the key insight that many crash states exhibit correlation in terms of crash consistency,  i.e., a set of crash states, $C_1$ can be implicitly tested by testing the consistency of another representative set of crash states $C_2$.

It is intractable to identify such a representative crash-state set as such a brute-force approach requires generating and exhaustively testing every crash state in an execution. 
Therefore, we propose an \emph{update behaviors-based} heuristic to \emph{approximate} this representative crash-state set. %

Our key observation is that \textbf{application developers often write code where semantically-related program updates are grouped together and make assumptions about the crash consistency of their programs at the granularity of these update behaviors.}

\add{
Therefore, multiple  update behaviors can expose the same crash-consistency bugs.
A \emph{representative} update behavior of a program can be tested on behalf of the other updates. 
If a crash-consistency bug is found due to the ordering constraints enforced (or not enforced) in the representative update, it is likely that the represented updates create crash states that are also crash-inconsistent (see \Cref{subsec:step-4} for the formal definition of representative testing in the context of crash-consistency testing).
}

We now describe two case studies on \posixBased and \mmioBased applications by first providing background for how these applications persist data to durable storage.
We then introduce two example crash-consistency bugs and demonstrate how the update behaviors-based heuristic can be applied to approximate the representative sets of crash states.

\subsection{Background}
\textbf{\posixBased applications.} 
Many applications are implemented on top of file systems that comply with the POSIX standard~\cite{posix}.
For persistence requirements, \posixBased applications interact with the underlying file system through system calls.
Therefore, the unit of an update behavior in \posixBased applications is a syscall.
To improve performance, the file system typically buffers application updates in a page cache, instead of directly flushing changes to the disk.
However, the POSIX standard is vague in terms of when the updates are persisted on disk~\cite{posixvague}.
Different file system implementations may have different behaviors, so a developer's misunderstanding about file system guarantees could easily lead to crash-consistency bugs~\cite{ALICE,torturingdb}.

For applications that need strong crash-consistency guarantees, developers use a combination of unit tests and stress tests to detect crash-consistency vulnerabilities~\cite{leveldb, rocksdb, wiredtiger}.
For example, RocksDB~\cite{wiredtiger} uses a suite of C++ unit tests.
While useful for regression testing and verification of bug fixes, these unit tests cannot find deep crash-consistency bugs in uncommon execution paths.
To compensate for this, RocksDB developers also run stress tests with a combination of write, deletion, compaction, reopen, and other operations.
However, exhaustively checking every possible crash state generated from stress tests is inefficient, so developers commonly
inject random crashes while running test workloads to test if the database is recoverable and consistent.

\textbf{\mmioBased applications.} 
Memory-mapped I/O (MMIO) is a method commonly used to perform communication between CPU and I/O devices through memory-mapped files.
MMIO provides a shared address space for both the main memory and I/O devices.
Developers can access devices through memory load and stores without the need for special I/O commands or system calls, improving the read/write performance.
As a result, the unit of an update behavior is a memory operation in \mmioBased applications.
MMIO has been adopted in many applications such as high-performance storage engines~\cite{lmdb, pebblesdb, hse} and persistent-memory (PM) libraries~\cite{pmdk,lee2019recipe, zuo2018level}.
Most \mmioBased applications need explicit memory ordering instructions (e.g. memory fence instructions, \stt{msync}) to enforce ordering between memory updates. 
Otherwise, the underlying implementation is free to order updates, which may further improve performance but leads to potentially more crash-consistency bugs due to increasing number of possible orderings~\cite{mahapatra2019dont, neal2020agamotto, neal2021hippocrates}.

\begin{figure}[t]
    \centering
    \begin{minipage}{.48\textwidth}
        \centering                \begin{lstlisting}[language=C,style=customnew,caption={Simplified snippet of a crash-consistency bug in \stt{SetCurrentFile} in RocksDB~\cite{rocksdb}. \stt{CURRENT} is not always synced and could be corrupted when a crash occurs.},label={lst:setcurrentfile_bug},escapechar=|]
// CURRENT file stores a pointer to the 
// latest MANIFEST meta-data file in 
// the database
IOStatus SetCurrentFile(..., 
    FSDir* dir_to_fsync) {
  Status s = WriteToNewFile(fs,ptr,tmp); |\label{line:tmp_write}|
  if (s.ok()) {
    TEST_KILL_RANDOM_WITH_WEIGHT(RAND); |\label{line:kill_before_rename}|
    s = fs->RenameFile(
        tmp, CurrentFileName(db)); |\label{line:rename}|
    TEST_KILL_RANDOM_WITH_WEIGHT(RAND); |\label{line:kill_after_rename}|
  }
  if (s.ok()) {
    if (dir_to_fsync != nullptr) { |\label{line:dir_to_fsync}|
      s = dir_to_fsync->SyncDir(
        CurrentFileName(db)); |\label{line:sync}|
    }
    |\hl{/* Crash occurs here! */}|
  } else { fs->DeleteFile(tmp); |\label{line:delete}|}
}
\end{lstlisting}
    \end{minipage}%
    \begin{minipage}{.48\textwidth}
    \centering
    \includegraphics[width=\linewidth]{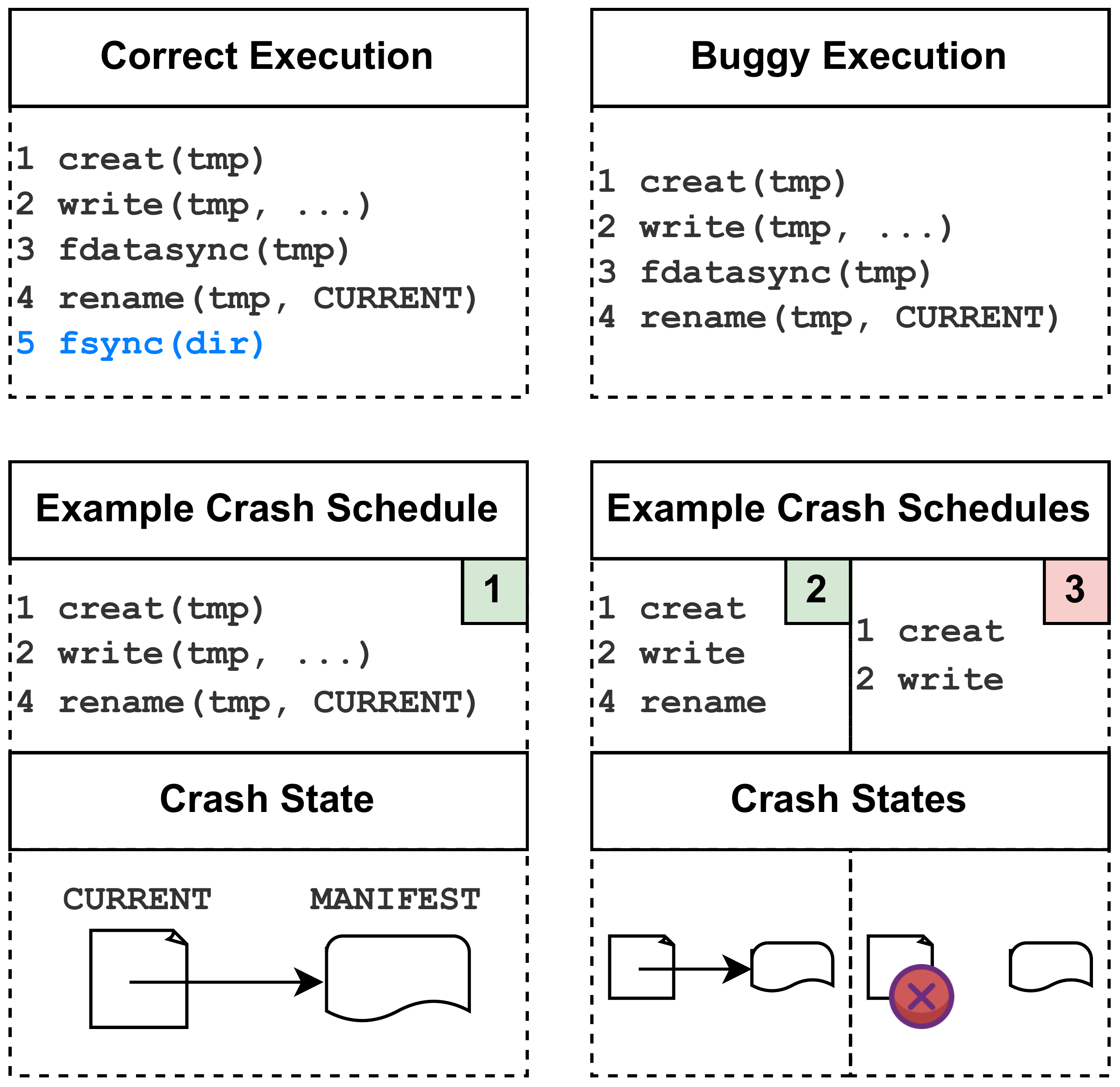}
    \caption{Correct execution trace and buggy execution trace of the \stt{SetCurrentFile} bug, with example crash schedules and resulting crash states from each execution.}
    \label{fig:posix_bug_example}
    \end{minipage}
    \vspace{-2em}
\end{figure}

\subsection{\posixBased Applications Bug Example}
\label{subsec:posix-case-study}

\Cref{lst:setcurrentfile_bug} shows an example of a crash-consistency bug in the \stt{SetCurrentFile} function reported by running a stress test in RocksDB~\cite{rocksdb}.
When a new key-value pair is inserted in RocksDB, it is first written to both a Write-Ahead Log (WAL) for durability and an in-memory MemTable for read performance.
When the MemTable is full, the system invokes background compaction to compact data into a compressed file.
Compaction will also update a \stt{MANIFEST} file which contains meta-data of all compressed files in the database.
The \stt{CURRENT} file keeps track of a pointer to the latest \stt{MANIFEST} file in the database.
When compaction completes, the program invokes \stt{SetCurrentFile} to update the pointer to \stt{MANIFEST}. Then it creates and writes to a temporary file first (\refline{line:tmp_write}) and renames it to be \stt{CURRENT} (\refline{line:rename}).
However, during execution, a rare code path may set \stt{dir\_to\_fsync} to be \stt{nullptr}, which skips the \stt{SyncDir} call to sync the directory (\refline{line:dir_to_fsync}).
Therefore, when a crash occurs after the rename, the un-synced directory containing the new \stt{CURRENT} file may encounter a data loss.
The stress test captures this bug by updating a global log file that keeps track of the expected state of the database, triggering an assertion failure when it detects an inconsistency.

\Cref{fig:posix_bug_example} shows a comparison of traces and example crash states between the correct and buggy execution.
\Cref{fig:posix_bug_example} also shows examples of sequences of operations that result in crash states (i.e., \emph{crash schedules}).
Without the \stt{fsync(dir)} call, the \stt{rename} operation may not be persisted to disk when \stt{SetCurrentFile} finishes, generating a buggy Crash State 3 for the buggy execution.

We also observe that although during each execution, the exact content written to \stt{tmp} file may be different, the orderings of syscalls in crash schedules that produce correct Crash State 1 and correct Crash State 2 are equivalent.

In this example, the buggy execution of \stt{Set\-Current\-File} function is the \emph{representative} update behavior.
This is because the sequence of operations in the buggy execution generates a superset of possible orderings of the sequence in the correct execution.
Therefore, it is sufficient to only test
the buggy execution to expose this crash-consistency bug.

\subsection{\mmioBased Applications Bug Example}
\label{subsec:mmio-case-study}

\begin{figure}[t]
    \centering
    \begin{minipage}{.48\textwidth}
        \centering                \begin{lstlisting}[language=C,style=customnew,caption={Simplified snippet of a crash-consistency bug in \stt{insert} in Level hashing~\cite{zuo2018level}. Untimely crashes in the \stt{insert} function lead to only the \stt{valid} flag being updated. A synthetic \stt{insert\_ordered} function produces a subset of orderings in \stt{insert}.},label={lst:level_insert_bug},escapechar=|]
// Assume entry spans multiple cache lines
typedef struct entry {
    uint8_t key[KEY_LEN], value[VALUE_LEN];
    bool valid;
} entry_t;  
void insert(key, value) {
    entry_t *new_entry = ...; |\label{line:insert_begin}|
    new_entry->key = key; 
    new_entry->value = value; 
    new_entry->valid = true; |\label{line:valid_set} \label{line:insert_end}|
    |\hl{/* Crash occurs here! */} \label{line:crash_site1}|
    PERSIST(new_entry); |\label{line:entry_persist}|
}
 void insert_ordered(key, value) { |\label{line:insert_key_only}|
    entry_t *new_entry = ...;
    new_entry->key = key;
    PERSIST(&new_entry->key); |\label{line:key_order}|
    new_entry->value = value;
    new_entry->valid = true;
    |\hl{/* Crash occurs here! */}|
    PERSIST(new_entry);
}
\end{lstlisting}
    \end{minipage}%
    \begin{minipage}{.48\textwidth}
    \centering
    \includegraphics[width=\linewidth]{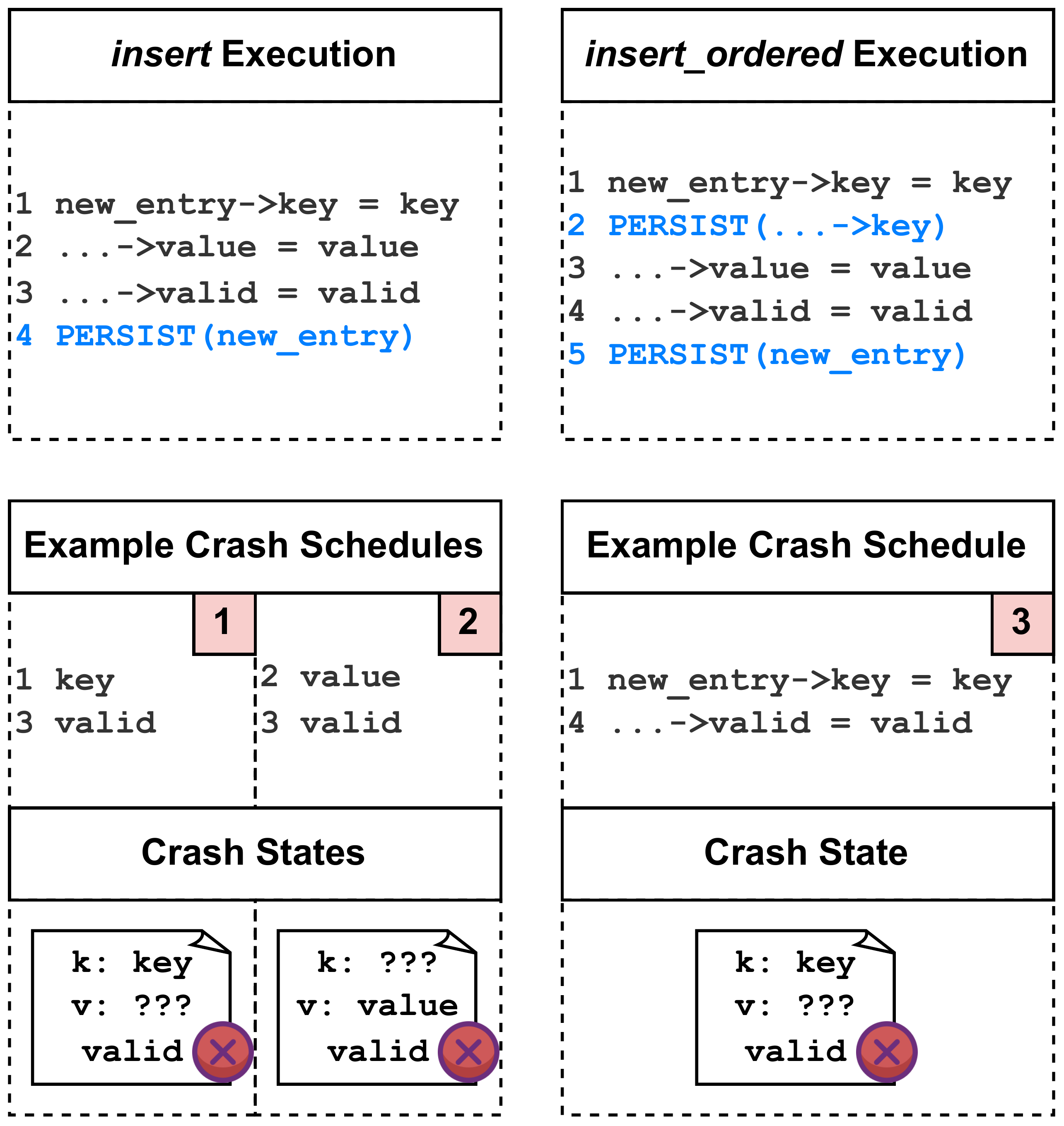}
    \caption{Executions from \stt{insert} and \stt{insert\_ordered} functions respectively, with example crash schedules and resulting crash states from each execution. }
    \label{fig:mmio_bug_example}
    \end{minipage}
    \vspace{-2em}
\end{figure}

 \Cref{lst:level_insert_bug} is a code snippet of a crash-consistency bug found in Level hashing~\cite{zuo2018level}.
In the \stt{insert} function, the application inserts a new key-value pair into a hash table entry and sets the \stt{valid} field to \stt{true} (lines \ref{line:insert_begin}--\ref{line:valid_set}). 
This update order would ensure crash consistency, but the updates to these fields are not guaranteed to be persisted due to the lack of \stt{PERSIST} calls and therefore may be persisted in any order.
If a crash occurs after the \stt{valid} field is set (\refline{line:valid_set}) but before the \stt{key} and \stt{value} are persisted (\refline{line:entry_persist}), the updates to \stt{key} and \stt{value} may be lost, resulting in inconsistent data and crash-consistency violations.

The synthetic \stt{insert\_ordered} function is an alternative implementation
of the \stt{insert} function that attempts to fix the aforementioned bug.
Since an additional ordering constraint is inserted between the \stt{key} and \stt{value}, the \stt{key} is guaranteed to be persisted before \stt{value} and \stt{valid}.
However, in this new \stt{insert\_ordered} function, a crash happening before the final \stt{PERSIST} call could still cause the same crash-consistency bug, since the \stt{valid} flag may still be persisted before \stt{value} and triggers the assertion failure.

We observe that model checking the possible orderings of updates to \stt{new\_entry} in \stt{insert} function and \stt{new\_entry} in \stt{insert\_ordered} function could expose the same crash-consistency bug.
\Cref{fig:mmio_bug_example} shows a comparison of example crash schedules and crash states resulted from \stt{insert} function execution and \stt{insert\_ordered} function execution respectively.
Crash States 1, 2 and 3 are all buggy since \stt{valid} flag is set and persisted before both \stt{key} and \stt{value}'s updates are persisted.
We also observe that although the exact content written to \stt{key} entry may be different, the orderings of MMIO updates in crash schedules that produce Crash State 1 and Crash State 3 are equivalent.

In this example, the set of updates to \stt{new\_entry} in \stt{insert} is the \emph{representative} update behavior.
This is because the sequence of updates to \stt{new\_entry} in \stt{insert} produces a superset of possible orderings of updates in \stt{insert\_ordered}, so only model checking the first sequence is sufficient to expose the bug.

In the following section, we will describe how to leverage our observations on representative update behaviors in \posixBased and \mmioBased applications to build a scalable and accurate crash-consistency testing tool.

\section{Design of \sys}
\label{sec:design}

\begin{figure*}[t]
    \centering
    \includegraphics[width=1.0\linewidth]{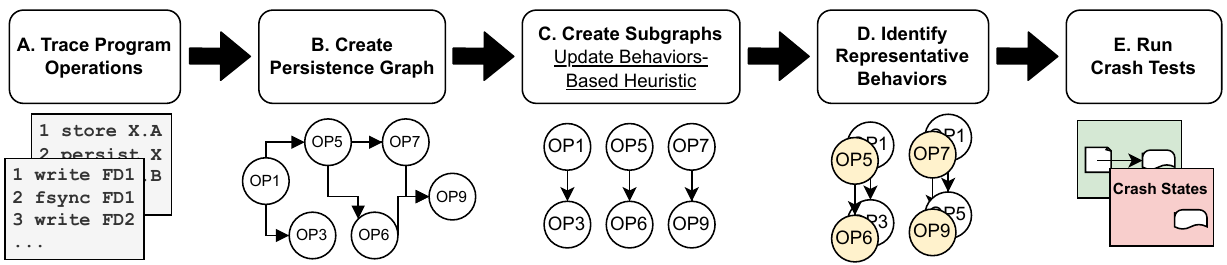}
    \caption{Functional overview of \sys.}
    \label{fig:overview}
    \vspace{-1em}
\end{figure*}

\sys{} implements representative testing by identifying \emph{update behaviors} in the program, grouping them together by their representative, and then testing the representative update behavior from each group for crash consistency.

\sys{} is based on crash-consistency model checking workflows (see \Cref{fig:overview}), which first execute an application to generate a trace of operations (i.e., syscalls in \posixBased applications; stores, flushes, and fences in \mmioBased applications).
\sys{} then performs crash-consistency testing by generating crash images that result from update orderings and querying an application-specific oracle to determine the consistency of each resulting image, while incorporating multiple optimizations as we detail below.

Specifically, \sys first traces program operations during execution (\Cref{fig:overview} Step A, see \Cref{subsec:step-1}). 
\sys then converts this trace into a \emph{Persistence Graph} that contains all updates in the execution as nodes and all orderings between updates as edges (\Cref{subsec:step-2} Step B).  
 \sys divides the graph into subgraphs using the \emph{update behaviors-based} heuristic (\Cref{subsec:step-3} Step C).
 The implementation of the heuristic highly depends on the characteristics of updates in the applications, and we elaborate the detailed algorithms used for \posixBased applications and \mmioBased applications in Section~\ref{sec:algorithms}.
 \sys then identifies representative subgraphs among all generated subgraphs (\Cref{subsec:step-4} Step D).
 Finally, \sys uses \emph{crash-consistency model checking} (\Cref{subsec:step-5} Step E) to test crash states from each representative subgraph.
 \sys leverages DPOR to avoid generating redundant crash states.
Crash-consistency model checking, as also described in prior work~\cite{ALICE, torturingdb, gorjiara2021jaaru,pmreorder,lantz2014yat}, validates crash states using a consistency-checking oracle (e.g., a specialized function or application recovery code).

In the remainder of this section, we discuss each step of \sys's testing process in detail (\Crefrange{subsec:step-1}{subsec:step-5}).
\add{
To clearly demonstrate the workflow of \sys, we will use two end-to-end running examples: one for POSIX-based applications and one for MMIO-based applications.
}

\subsection{Tracing Program Operations (Step A)}
\label{subsec:step-1}

\sys traces program operations during an execution of the application under test.
\sys can then reason about all of the possible persistence orderings over operations in the trace, since the trace describes all dependencies between program operations (\cref{subsec:step-2}). 
For \posixBased applications, the trace contains a list of syscalls (e.g. write, rename, unlink, fdatasync, and fsync). 
For \mmioBased applications, the trace includes a list of memory update operations (i.e., stores, cache-line flushes, and memory fences; \cref{subsec:mmio-case-study}).
The operations in the trace are generated in the order in which they are executed by the CPU.

\begin{figure}[t]
    \centering
    \includegraphics[width=0.95\linewidth]{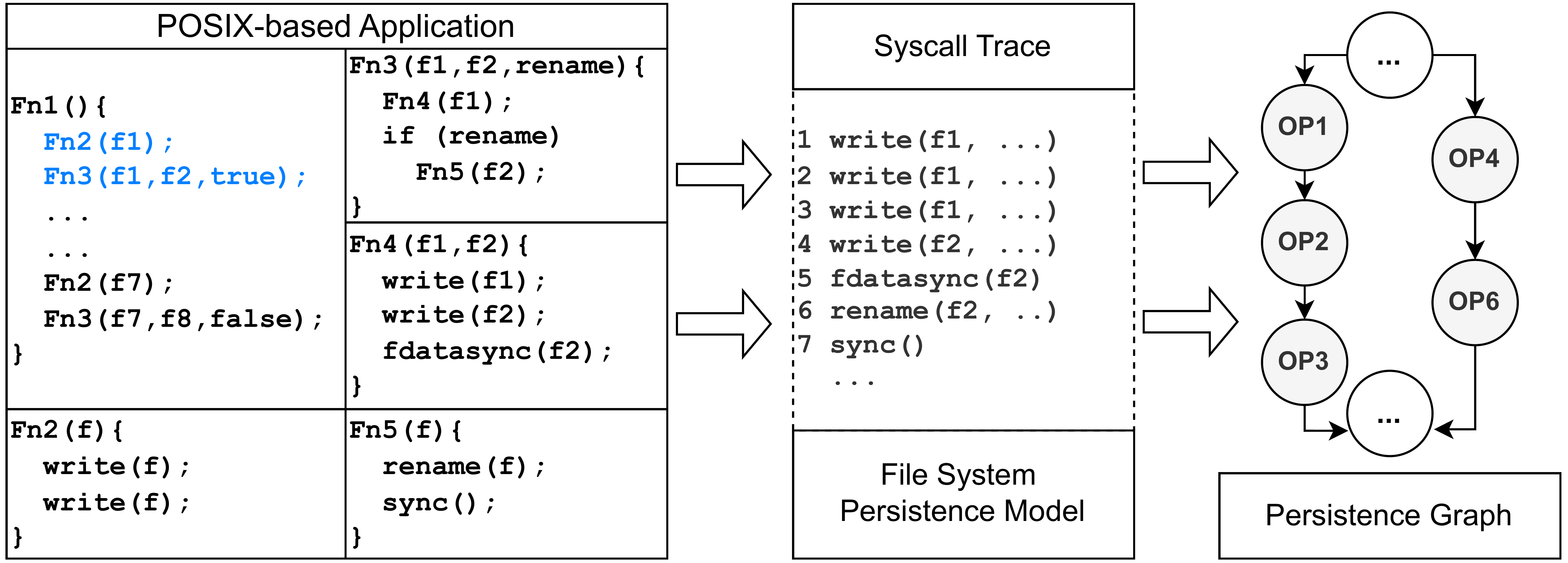}
    \caption{Persistence graph creation for an example POSIX-based application (Step A \& B). \texttt{Fn1} is the function currently under execution, which invokes functions \texttt{Fn2} to \texttt{Fn5}. Each function may persist data to the disk or alter file status on the disk.
    }
    \label{fig:persistence-graph-posix}
\end{figure}

\begin{figure}[t]
    \centering
    \includegraphics[width=0.95\linewidth]{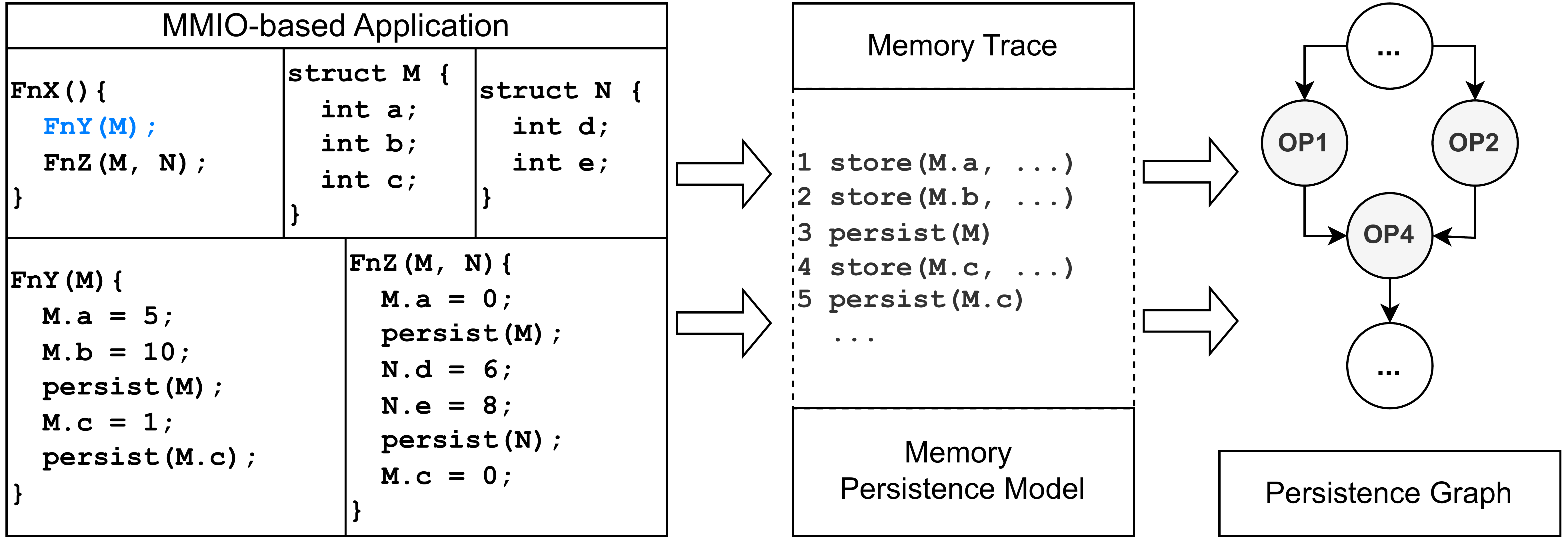}
    \caption{Persistence graph creation for an example MMIO-based application (Step A \& B). \texttt{FnX} is the function currently under execution, which calls functions \texttt{FnY} and \texttt{FnZ} to persist data in data structure \texttt{M} and \texttt{N} to disk.
    }
    \label{fig:persistence-graph-mmio}
\end{figure}

\Cref{fig:persistence-graph-posix} shows an example POSIX-based application that persists data to the disk via syscalls.
Function \texttt{Fn1} is the main function that is under execution.
\textit{Syscall Trace} gives an example of syscalls recorded after executing the functions colored in blue in function \texttt{Fn1}. 
Similarly, \Cref{fig:persistence-graph-mmio} is an example MMIO-based application that persists data via memory update operations.
\textit{Memory Trace} records memory operations after executing the functions colored in blue in function \texttt{FnX}.
Different program executions may generate traces that perform different updates (i.e., different paths through the program); we discuss trace coverage in \cref{sec:discussion}.

\subsection{Persistence Graph Construction (Step B)}
\label{subsec:step-2}

\sys provides a \emph{Persistence Graph} abstraction to represent all the operations in a program execution trace, along with the "happens-before" dependencies for persistence between each operation~\cite{Ferrite, x86memorymodel}.
A persistence graph can be generated using an execution trace (from Step A) and a persistence model.
A persistence model describes how sequences of operations persist data.
The persistence graph represents each operation in the trace as a node, and represents the "happens-before" dependencies as edges.

\Cref{fig:persistence-graph-posix} shows the example persistence graph generated by \sys for \posixBased applications.
The \textit{Syscall Trace} from the previous step and a \textit{File System Persistence Model} are used to generate the persistence graph, since different file systems' implementation and configuration may affect how the syscalls can be reordered \cite{ALICE}.
\sys currently uses the ext4 file system model proposed by Ferrite \cite{Ferrite}, but it could be adapted to test \posixBased applications running on other file systems by defining new file system models. 
In the example, assuming the three write syscalls to  file $f1$ writes to the same block (operations 1, 2 and 3), the first write must be persisted before the second and the second write must be persisted before the third.
The fdatasync on file $f2$ (operation 5) guarantees that the write to $f2$ (operation 4) will be persisted before the rename of $f2$ (operation 6).
There is a global sync at the end (operation 7) so the nodes have outgoing edges to subsequent nodes in the graph.

\add{
Similarly, \Cref{fig:persistence-graph-mmio} shows the example persistence graph for \mmioBased applications.
A \textit{Memory Trace} and a \textit{Memory Persistence Model} are provided to generate a persistence graph.
\sys currently implements the "persistent x86" memory persistence model proposed by prior work \cite{x86memorymodel}.
}

\subsection{Subgraph Creation (Step C)}
\label{subsec:step-3}

After constructing the persistence graph (Step B), \sys splits the graph into subgraphs to identify the program's update behaviors.

\sys creates one subgraph from the persistence graph for each update behavior in the program.
Based on our case studies in \cref{subsec:posix-case-study} and \cref{subsec:mmio-case-study}, \sys leverages an \emph{update behaviors-based} heuristic to automatically infer and group update behaviors in an execution.
The implementation of the heuristic is dependent on characteristics of updates in different types of applications (see \cref{sec:algorithms}).

\subsection{Grouping Update Behaviors via Representatives (Step D)}
\label{subsec:step-4}
\label{subsec:grouping}

After identifying the program's update behaviors (Step C), \sys groups behaviors by their representatives.
\sys considers one behavior to represent another if the behavior contains a superset of the program updates and enforces a subset of the dependencies demonstrated by the represented behavior.
By this definition, the representative behavior will contain a superset of possible crash schedules and therefore create a superset of equivalent crash-states.

Below, we formalize the \textit{represents} relation and define its underlying components:

\begin{definition}[Update Behavior]
\label{def:update-behavior}
    An \textit{update behavior} $U$ is a subset of a program's execution that persists data to durable storage, such as a single function call that issues several syscalls to update a file or a set of function calls that performs MMIO.
    As an update behavior includes storage operations, it may be represented by a \textit{persistence graph} ($P(U)$).
\end{definition}

\begin{definition}[Persistence Graph]
\label{def:persistence-graph}
    A persistence graph ($P$, or $P(U)$ where $U$ is an update behavior) is a directed, acyclic graph. A persistence graph is composed of persistence graph \textit{nodes} and \textit{edges}, representing storage updates that occur during a program's execution and their corresponding happens-before dependencies. 

    \textbf{Nodes.}
    The \textit{nodes} of a persistence graph ($N(P)$) are operations that update durable storage, such as a \texttt{write} syscall or a \texttt{memcpy} to an MMIO address. 
    A node $n$ in a program's persistence graph is created by the runtime execution of a storage operation ($Op(n)$) at a particular application binary location (e.g., when a \texttt{write} syscall is executed from an invocation of \texttt{update\_db()}).
    Persistence graph nodes are therefore composed of \textit{static information} (source code/application binary location, or $Loc(n)$) and \textit{dynamic information} (call stack of the operation, values of arguments provided to invocation site at runtime).

    \textbf{Edges.}
    The \textit{edges} in a persistence graph ($E(P)$) are happens-before dependencies between storage operations that are enforced by the program, the CPU, or the underlying file system. 
    For example, an \texttt{fsync} call between two \texttt{write} syscalls to the same file (i.e., the sequence \texttt{write(f1, ...); fsync(f1); write(f1, ...)}) enforces a \texttt{dependency} between the two \texttt{write}s and therefore creates a directed edge from the first \texttt{write} to the second \texttt{write} in the persistence graph.
    When considering possible durability orderings, update operations with dependencies must be made durable in an order that respects the happens-before relationship represented by the dependency to ensure data consistency.
    Updates without an edge in the persistence graph between them have no happens-before relationship and therefore may 
    be persisted in any order in relation to one another.
    
\end{definition}

We also define the set of edges over an arbitrary set of nodes $E(N_a)$ as equal to $E(P)$ if $N_a = N(P)$, and otherwise define $E(N_a) = \{e \in E(P) | \exists a, b \in N_a \text{ such that } e = (a, b)\}$.

\begin{definition}[Persistence Graph Node Equivalence]
\label{def:persistence-node-equivalence}

    \add{
    We define two persistence graph nodes ($n_1$, $n_2$) as \textit{equivalent} ($n_1 \sim n_2$) if they contain the same \textit{static information} (see \Cref{def:persistence-graph}).
    Equivalent nodes are created from the execution of the same program location, and will therefore perform the same operation (i.e., if $n_1 \sim n_2$, then $Op(n_1) = Op(n_2)$), but potentially with different \textit{dynamic information} such as different call stacks or runtime arguments.
    The node equivalence is commutative (i.e. $n_1 \sim n_2 \Leftrightarrow n_2 \sim n_1$).
    This definition of equivalence is derived from our observations about how update behaviors are structured (\cref{subsec:general-methodology}) and enable us to define the \textit{represents} relation below (\Cref{def:represent}).
    }

    \add{
    We further define the notion of \textit{subset equivalence} ($\subseteq_\sim$) on two sets of nodes, $N_1$ and $N_2$, as:
    $$ N_2 \subseteq_\sim N_1 \Leftrightarrow \forall a\in N_2, \exists b \in N_1 \text{ such that } a \sim b $$

    We also define the \textit{equivalence image} of $N_1$ on $N_2$ ($N_{1\sim} (N_2)$) to be the subset of $N_1$ that is equivalent to elements in $N_2$, meaning:
    $$ N_{1\sim}(N_2) = \{ n \in N_1 | \exists a \in N_2, n \sim a\}$$

    From this definition, it follows that $ N_2 \subseteq_\sim N_1 \implies |N_{1\sim}(N_2)| = |N_2|$.
    }
\end{definition}

\begin{definition}[Persistence Graph Edge Equivalence]
\label{def:persistence-edge-equivalence}
    Similar to the equivalence of persistence graph nodes (\cref{def:persistence-node-equivalence}), we define equivalence between two edges ($e_1$, $e_2$) in persistence graphs as a function of the equivalence of the nodes the edges connect.
    As persistence graphs are directed graphs, a persistence edge $e$ can be represented as an ordered pair $(n_{s}, n_d)$, where $n_s$ is the source node and $n_d$ is the destination node.
    We then define two edges $e_1$ and $e_2$ as equivalent ($e_1 \sim e_2$) as the following:
    $$e_1 \sim e_2 \Leftrightarrow n_{s_1} \sim n_{s_2} \text{ and } n_{d_1} \sim n_{d_2}, \text{where } e_1 = (n_{s_1}, n_{d_1}) \text{ and } e_2 = (n_{s_2}, n_{d_2})$$

    We further define the notion of \textit{subset equivalence} ($\subseteq_\sim$) on two sets of edges, $E_1$ and $E_2$, as:
    $$ E_1 \subseteq_\sim E_2 \Longleftrightarrow \forall a\in E_1, \exists b \in E_2 \text{ such that } a \sim b $$
    
\end{definition}

\begin{definition}[Represents Relation]
\label{def:represent}
    Finally, we define the \textit{represents} relation $\sim_r$ over two update behaviors $U_1$ and $U_2$. 
    We say that $U_1$ \emph{represents} $U_2$ (i.e., $U_1 \sim_r U_2$) for the persistence graphs of the update behaviors $P(U_1)$ and $P(U_2)$:
    $$ N_1 := N(P(U_1)), N_2 := N(P(U_2)), N_{1\ image} := N_{1\sim}(N_2)$$
    $$U_1 \sim_r U_2 \Leftrightarrow P(U_1) \sim_r P(U_2) \Leftrightarrow N_2 \subseteq_\sim N_1 \text{ and } E(N_{1\ image}) \subseteq_\sim E(N_2)$$
    
    Informally, this definition means that $U_2$ contains a subset of nodes in $U_1$ in the persistence graph ($N_2 \subseteq_\sim N_1$), and the subset of nodes in $U_1$ that are equivalent to nodes in $U_2$ ($N_{1\ image}$) have the same or fewer dependencies as those nodes in $U_2$ (i.e., $E(N_{1\ image}) \subseteq_\sim E(N_2)$).
\end{definition}

By \cref{def:represent}, a representative in the group $G$ is the update behavior $U$ such that $\forall U' \in G, U \sim_r U'$. 
Furthermore, $U_1$ and $U_2$ are in the same group $G$ if $U_1 \sim_r U_2$ or $U_2 \sim_r U_1$.

\begin{figure}[t]
    \centering
    \includegraphics[width=0.7\linewidth]{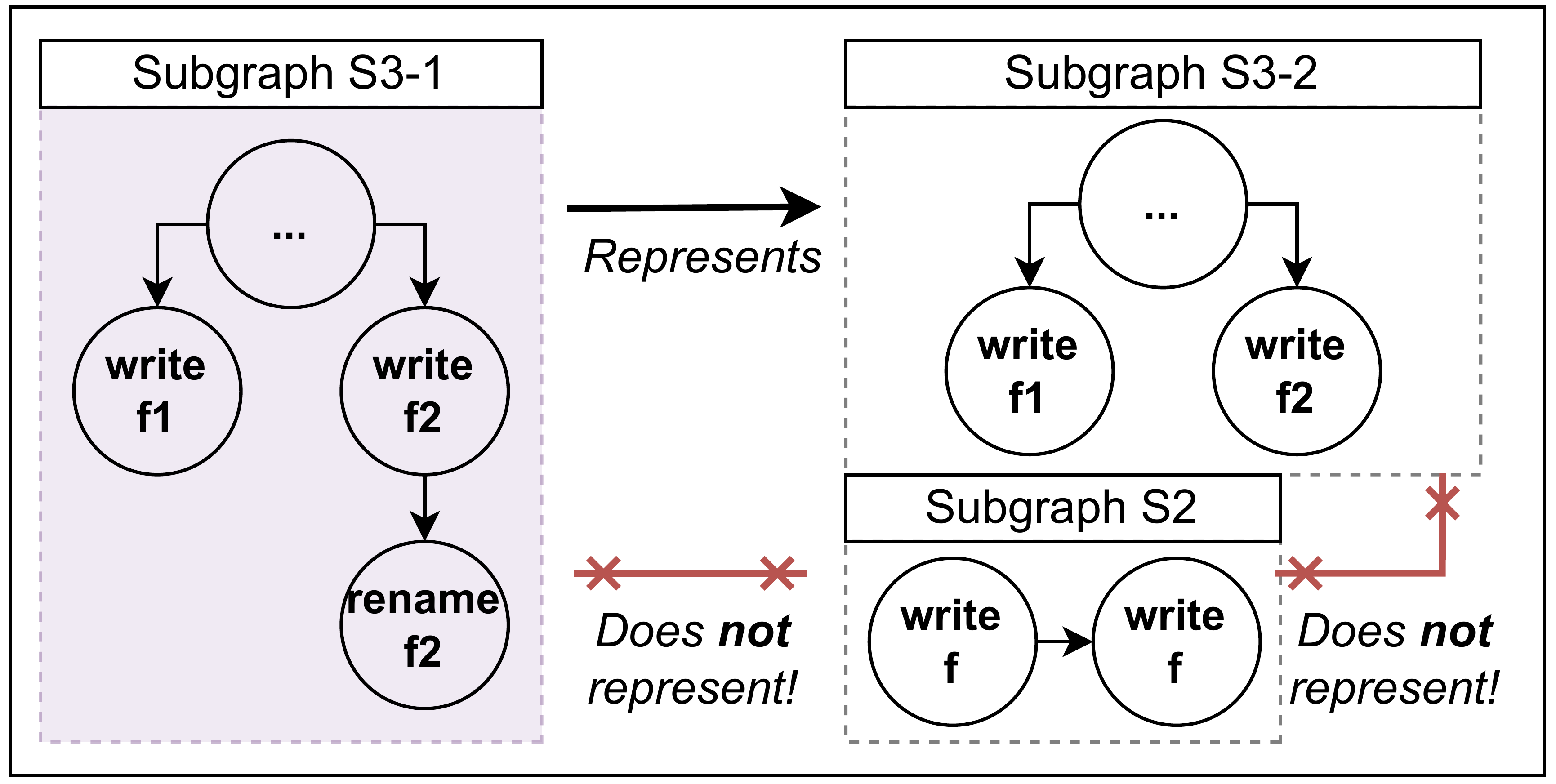}
    \caption{An example of representative relations between subgraphs created from the \posixBased application in \Cref{fig:persistence-graph-posix}.}
    \label{fig:similarity}
\end{figure}

\cref{fig:similarity} provides an example of how the subgraphs created from the example \posixBased application in \Cref{fig:persistence-graph-posix} are placed into a group.
For function \texttt{Fn3} of the example in \Cref{fig:persistence-graph-posix}, there are two possible types of subgraphs: Subgraph S3-1 and Subgraph S3-2, depending on whether the \texttt{rename} flag is enabled.
Subgraph S3-1 represents Subgraph S3-2 because Subgraph S3-2 contains a subset of the nodes in Subgraph S3-1, $\{write(f1), write(f2)\}$ $\subset$ $\{write(f1), write(f2), rename(f2)\}$ and the dependencies in Subgraph S3-2 are the same as the dependencies among equivalent nodes in Subgraph S3-1.
However, Subgraph S2 created from executing function \texttt{Fn2} does not have representative relations with either Subgraph S3-1 or Subgraph S3-2, since Subgraph S2 contains a different set of nodes.

\sys identifies representative update behaviors as follows. 
\sys iterates through all identified update behaviors in sorted order, from largest update behavior to smallest. 
It adds each update behavior to every existing group that represents it.
If there is no group that represents the update behavior, then \sys forms a new group for the update behavior.
When multiple subgraphs could be chosen as the representative, \sys chooses any one of the subgraphs to be representative.
In \cref{fig:similarity}, if we assume that Subgraphs S3-1, S3-2, and S2 are the only subgraphs, since Subgraph S3-1 represents Subgraph S3-2, S3-1 would be chosen as a representative update behavior for the group consisting of Subgraphs S3-1 and S3-2.
Since neither Subgraph S3-1 nor S3-2 represents Subgraph S2 and Subgraph S2 does not represent either Subgraph S3-1 or S3-2, S2 would be its own representative.

\subsection{Model Checking (Step E)}
\label{subsec:step-5}

Finally, \sys tests the representative update behavior for each group using a model checker.
For each representative, the model checker enumerates all sequences over subsets of operations from the persistence graph in Step B (\cref{subsec:step-2}), ignoring sequences pruned by DPOR. Each operation sequence is replayed to get a resulting crash state. Operations that occur in the application trace before the representative update behavior are also made persistent.
For each crash state, \sys runs the application in recovery mode and checks for inconsistencies in the persistent data with an application-defined oracle.

\section{Update Behavior Derivation Algorithms}
\label{sec:algorithms}

 We now detail the algorithms for update behavior derivation for both \posixBased and \mmioBased applications.
In order to determine representatives,
 \sys needs to first identify which operations are associated with an update behavior.
 Depending on the update behavior characteristics described in Section~\ref{sec:rep-testing}, \sys has different implementations for the \emph{update behaviors-based heuristic.}
For \posixBased applications, \sys{} uses the backtrace (i.e. the call stack with function name, source code file and line number) provided in the trace to derive update behaviors, grouping update behaviors by traversing the call stack tree to generate more possible crash states to test.
For \mmioBased applications, \sys{} employs a heuristic approach to determine 
if MMIO operations are related based on (1) the data types and instances modified, and (2) the temporal locality of the MMIO operations.

\subsection{\posixBased Applications}

\begin{algorithm}[t]
  \algrule{}

  \KwIn{Persistence Graph $\mathcal{G}$,
        Trace $\mathcal{T}$
  }
  \KwOut{Subgraph Map $\mathcal{M}(f \rightarrow L)$: Function Name $f$ to List of Subgraphs $L$}

  \algrule{}

  \SetKwProg{Function}{Function}{:}{}
  \Function{$\mathrm{Get\_POSIX\_Subgraph}$($\mathcal{G}$, $\mathcal{T}$)}{

    \Comment{Split trace by thread and initialize longest common prefix of backtraces for previous and current iteration}
    
    $\mathcal{T} \leftarrow \{\mathcal{T}[tid] \mid tid \textrm{ in } \mathcal{T} \}$, $prev\_lcp, lcp \leftarrow \emptyset$\; \label{algo_line:split_by_tid}

    \For{\upshape every thread ID $tid$} {
        
        $ \{op_i, op_{i+1}\} \leftarrow \mathcal{T}[tid]$\;
        
        \Comment{Get longest common prefix of backtraces}
        
        $ lcp \leftarrow \textrm{Get\_LCP}(op_i, op_{i+1}) $\; \label{algo_line:get_lcp}

      \If{\upshape $op_i$ is not in any update behavior}{
        \Comment{Create a new update behavior}
        
        $ub \leftarrow \{op_i, op_{i+1}\}$\; \label{algo_line:create_ub}
      }
      \Else{
        \Comment{Get the functions that longest common prefixes each ends at}

        $ prev\_f, f \leftarrow \textrm{Get\_Func}(prev\_lcp, lcp)$\;
        
        \uIf{\upshape $prev\_f = f$ }{ \label{algo_line:lcp_same_depth}

            $ub \leftarrow ub \cup \{ op_{i+1} \}$\;
            
        }
        \uElseIf{\upshape $prev\_f$ is parent of $f$}{
        \label{algo_line:lcp_deeper}
            \Comment{Remove $op_i$ from $ub$, end current $ub$, create a subgraph, and create new $ub$}
            
            $ub \leftarrow ub \setminus \{ op_{i} \}$, $\mathcal{G}_{ub} \leftarrow \mathcal{G}[ub]$\;

            $\mathcal{M}(prev\_f) \leftarrow \mathcal{M}(prev\_f) \cup \{ \mathcal{G}_{ub}$ \} \;

            $ub = \{op_i, op_{i+1}\}$\;
        } 
        \uElse{
        \label{algo_line:lcp_shallower}
            \Comment{End current $ub$ and create a subgraph}
            $\mathcal{G}_{ub} \leftarrow \mathcal{G}[ub]$\;

            $\mathcal{M}(prev\_f) \leftarrow \mathcal{M}(prev\_f) \cup \{ \mathcal{G}_{ub}$ \} \;
        }
      }
      $prev\_lcp \leftarrow lcp$, $i \leftarrow i+1$\;
    }
    \Return{$\mathcal{M}$}
    }
    
  \algrule{}
  \caption{Get\_POSIX\_Subgraph: Create POSIX subgraph by deriving update behaviors based on backtrace of adjacent operations (\cref{subsubsec:C1-POSIX}).}
  \label{algo:get_posix_subgraph}
\end{algorithm}

\paragraph{Functions-related update behaviors}
From \Cref{subsec:posix-case-study}, we observe that the crash consistency of \posixBased applications commonly relies on the consistency of individual functions.
In \Cref{lst:setcurrentfile_bug}, on lines \ref{line:tmp_write}, \ref{line:rename} and \ref{line:sync}, all of the syscalls in the \stt{SetCurrentFile} function are effectively modifying the same file and should be executed atomically.
It is sufficient to test the \emph{representative} update behavior of the function (i.e., the buggy execution in this case), which covers all possible orderings generated by other update behaviors (i.e., the individual calls on lines \ref{line:tmp_write}, \ref{line:rename} and \ref{line:sync}).
However, requiring the developers to provide annotations for these functions adds a high burden for testing and may still miss deep crash-consistency bugs exposed from rare execution paths.
For example, lines~\ref{line:kill_before_rename} and \ref{line:kill_after_rename} in \Cref{lst:setcurrentfile_bug} show RocksDB developers' attempt to inject crash points before \stt{RenameFile} and after \stt{RenameFile} to detect potential crash-consistency vulnerabilities.
Despite the efforts, it still misses the bug explained above.
Instead, \sys{} uses a backtrace-based algorithm to automatically identify the update behaviors from the execution trace, enabling it to test more execution paths.

\paragraph{Grouping operations} Determining the granularity of update behaviors presents a trade-off between efficiency and coverage.
All syscalls could be trivially grouped into a single update behavior under \stt{main} function, but this would lead \sys to enumerate all possible orderings in an execution.
Grouping a syscall to the function that directly invokes it could generate far fewer crash states, but may miss important orderings that are only observable from functions in the call stack.

\sys{} achieves a middle ground between these two extremes using a two-step approach.
\sys{} first groups syscalls based on functions that are commonly observed between syscalls in the backtraces and generates function-specific subgraphs from the persistence graph (\cref{subsubsec:C1-POSIX}).
\sys{} then merges function subgraphs when possible, constructing update behaviors based on functions in the call stack (\cref{subsubsec:C2-POSIX}).

\subsubsection{Function Subgraph Creation}
\label{subsubsec:C1-POSIX}

\quad

In order to derive update behaviors, \sys constructs \emph{function subgraphs} based on the backtrace of issued operations. Specifically, \sys groups together all operations that share the same backtrace. Each function subgraph includes the nodes corresponding to those operations and all edges between those nodes from the original persistence graph. We refer to these function subgraphs as update behaviors, though they may be further combined as \cref{subsubsec:C2-POSIX} describes.

Algorithm~\ref{algo:get_posix_subgraph} shows the pseudocode for grouping syscalls to the function that can be commonly observed between syscalls in the backtrace.
The algorithm starts by first splitting the operations in the trace by thread ID (line~\ref{algo_line:split_by_tid}).
Since adjacent operations on different threads may have a majority of non-overlapping backtraces, only deriving update behaviors based on per-thread operations is meaningful.
On each thread, we iteratively calculate the longest common prefix of backtraces between pairs of adjacent operations (line~\ref{algo_line:get_lcp}).
The longest common prefix of backtraces is defined as the common function deepest in the call stack between pair of operations, where all parent stack frames are exactly the same.
If the current operation is not in any update behavior, we create a new one for the pair of operations in the current iteration (line~\ref{algo_line:create_ub}).
Otherwise, we compare the longest common prefixes between previous iteration and the current iteration.

If the longest common prefixes are the same between two iterations (line~\ref{algo_line:lcp_same_depth}), this indicates that the current update behavior is still active and we continue to group operations into it.
If the longest common prefix of previous iteration is shorter than the current iteration (line~\ref{algo_line:lcp_deeper}), this means that a new function is invoked in the execution and the current pair of operations should be grouped to the new function.
Therefore, we remove $op_i$ from the current update behavior, end the current update behavior, create a subgraph for it from the whole persistence graph and create a new update behavior for the pair of operations under the new function.
Finally, if the longest common prefix of previous iteration is longer than the current iteration (line~\ref{algo_line:lcp_shallower}), this means the current function has ended.
We again end the current update behavior and generate a new subgraph for it.

\subsubsection{Function Subgraph Grouping}
\label{subsubsec:C2-POSIX}

\quad

\begin{figure}[t]
    \centering
    \includegraphics[width=0.8\linewidth]{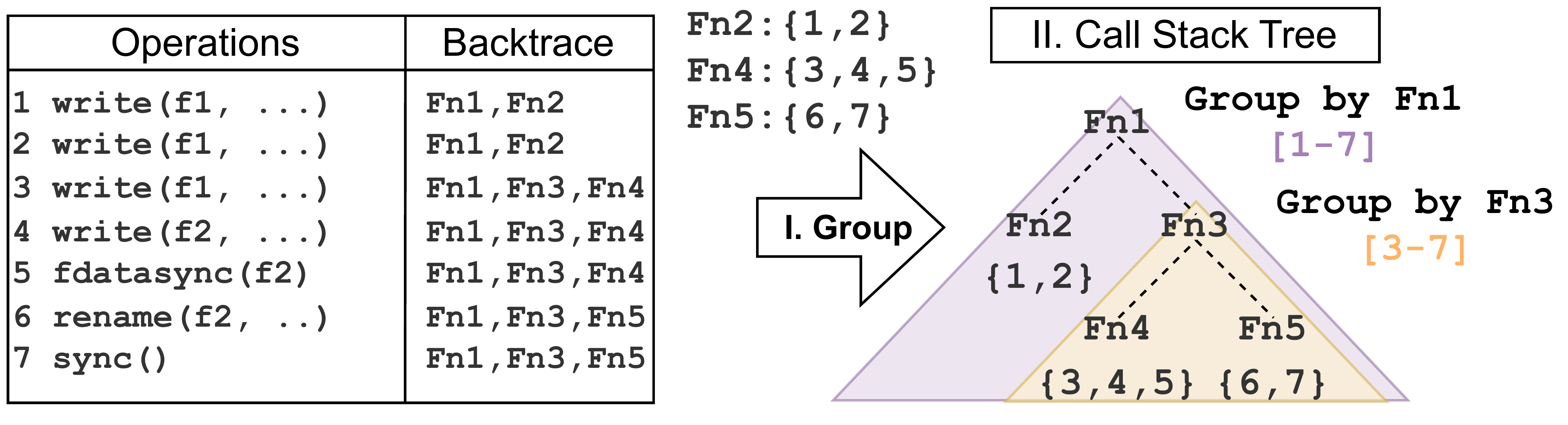}
    \caption{Call stack tree grouping example (\cref{subsubsec:C2-POSIX}).}
    \label{fig:stack_call_tree}
    \vspace{-1em}
\end{figure}

Since only testing update behaviors from individual functions may miss bugs, \sys{} iteratively combines update behaviors between parent and child functions in the call stack.
\sys{} maintains a call stack tree that expresses all parent-child relations from backtrace of operations in the trace to allow easier grouping.
\add{
\Cref{fig:stack_call_tree} shows an example of update behavior grouping based on the call stack tree for the example \posixBased application in \Cref{fig:persistence-graph-posix}.
Each node in the call stack tree represents a function, with a list of update behaviors grouped under it.
Function \texttt{Fn1} is the outermost function under execution.
\sys{} follows the call stack tree from leaf nodes to the root iteratively to group subgraphs constructed in \cref{subsubsec:C2-POSIX}.
For example, the update behavior [3--7] 
is grouped under function \texttt{Fn3} by merging the subgraphs from \texttt{Fn4} and \texttt{Fn5}.
Similarly, the update behavior [1--7] under \texttt{Fn1} is merged from the graphs under \texttt{Fn2} and \texttt{Fn3}.
}

\paragraph{Grouping temporally-located behaviors} For some executions, there may be multiple update behaviors under the same function that are temporally dispersed.
For example, function \emph{C} may be a background call that is invoked periodically. 
Na\"ively grouping all update behaviors in \emph{C} under function \emph{A} will create an update behavior that encapsulates unrelated operations, causing \sys to test redundant crash states.
Instead, after grouping all update behaviors from child function to the parent function, \sys runs a clustering algorithm to split the resulting update behavior into a list of smaller, temporally located update behaviors.
\sys uses the DBSCAN~\cite{DBSCAN} clustering algorithm, which can find behavior clusters of arbitrary shape and size.

\subsection{\mmioBased Applications}

\paragraph{Data structures-related update behaviors}
From our findings in \Cref{subsec:mmio-case-study},  we observe that many \mmioBased applications have \emph{data structures-related update behaviors}---updates to single data structures with high temporal locality.
In \Cref{lst:level_insert_bug}, the crash-consistency violation that occurs is dependent on a single data structure---\stt{entry\_t}.
Furthermore, the set of updates that causes the \stt{entry\_t} to become inconsistent occur solely within the context of the \stt{insert} or \stt{insert\_ordered} function, regardless of the history of other updates that may or may not have happened to that \stt{entry\_t} instance.
As an example, to test the crash consistency of the \stt{entry\_t} data structure, it is sufficient to test the \emph{representative} update behavior of \stt{entry\_t} in \stt{insert}, which covers all possible orderings generated by update behavior of \stt{entry\_t} in \stt{insert\_ordered}.

\sys uses the aforementioned observations to build a heuristic method to identify the update behaviors of \mmioBased applications in the execution trace.
First, \sys groups updates together based on data type and instance and creates \textit{instance subgraphs} for each instance \sys has encountered (\cref{subsubsec:C1-MMIO}).
Then, \sys refines these groups based on temporal locality, by splitting the instance subgraphs into \textit{epoch subgraphs}, creating update behaviors that are suitable to apply representative testing (\cref{subsubsec:C2-MMIO}).

\smallskip

\subsubsection{Type Subgraph Creation} 
\label{subsubsec:C1-MMIO}

\quad

\begin{figure}[t]
    \centering
    \includegraphics[width=0.7\linewidth]{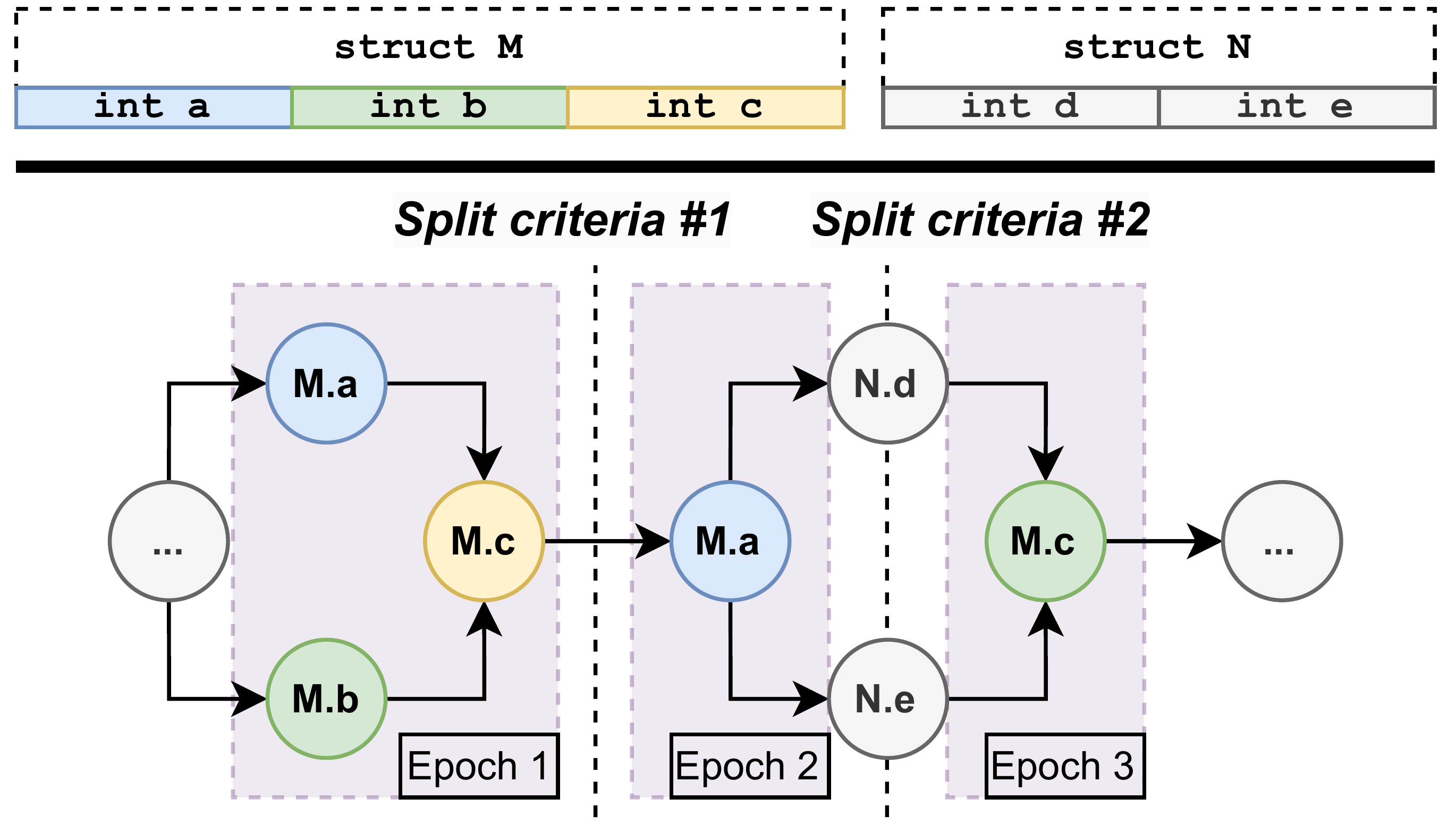}
    \caption{An example of how the persistence graph is grouped into type subgraphs and split into epoch graphs (\cref{subsubsec:C1-MMIO} \& \cref{subsubsec:C2-MMIO}).}
    \label{fig:types_and_epochs}
    \vspace{-1em}
\end{figure}

Motivated by our observation that \mmioBased applications typically place related data in a single data structure, \sys splits the persistence graph into subgraphs by data structure and instances.
\sys creates a \emph{type subgraph} for each data structure.
A type subgraph is a subgraph of the persistence graph obtained from \cref{subsec:step-2}, which includes all nodes and dependencies from the persistence graph that affect the relevant type (i.e. all memory operations associated with the data type and all corresponding "happens-before" relations).
\sys also creates subgraphs for a composite graph.
\sys captures ordering requirements among the constituent types of a composite type (e.g., a struct that has another struct as a field) since constituent types in a composite type can have crash-consistency ordering requirements.

\sys further splits each type subgraph into \textit{instance subgraphs}, since the inividual object is the unit of crash consistency. Each node in the type subgraph refers to a data field from a specific instance through the operation that the node represents.
\sys creates instance subgraphs for all instances of the data type by assigning each node in the type subgraph to the instance subgraph it belongs to based on the object referred by the node.
One subgraph is constructed for each instance of the pertinent data type, and contains only nodes and dependencies whose corresponding operation affects the respective instance.
\add{
\cref{fig:types_and_epochs} shows an example of updates for two data types $M$ and $N$, where the persistence graph is generated from the example \mmioBased application in \Cref{fig:persistence-graph-mmio}.
Data structure $M$ has three integer data fields: $M.a$, $M.b$ and $M.c$.
The 5 nodes in purple shaded region along with dependencies between them constitute an instance subgraph for data structure $M$.
}

\subsubsection{Epoch Subgraph Creation} 
\label{subsubsec:C2-MMIO}

\quad

\sys further refines instance subgraphs into update behaviors based on temporal locality of operations. Specifically, we cluster nodes into \emph{epochs}, and decompose each instance subgraph into many \emph{epoch subgraphs}.

The key operation that \sys is looking for is the \stt{PERSIST} operation between different updates to the same data structure in an MMIO-based application, as it may indicate the end of an application-specific task, thus the end of an update behavior.

\sys splits the instance subgraph into epochs before any update operation $I$ when either of the following criteria are met:

\begin{enumerate}
    \item All updates to the instance issued before $I$ are persisted \emph{and} $I$ updates a previously updated field of the instance.

    \item All updates to a different instance (that $I$ does not operate on) that are issued before $I$ are persisted before $I$.
\end{enumerate}

The resulting epoch subgraphs represent the  update behaviors in the execution.
Once an epoch is created this way, the tracking for repeated fields resets at the beginning of the new epoch.

\add{
\cref{fig:types_and_epochs} shows an example where both criteria lead \sys to create a new epoch. The first criteria demarcates Epoch 1 from Epoch 2, since the execution again updates $M.a$ after it persists all prior updates to $M$ $(\{M.a, M.b, M.c\})$. Epochs 2 and 3 are divided due to the second criteria.
Here, the updates to $N.d$ and $N.e$ are persisted between two updates to $M$ ($M.a$ and $M.c$).
Since $M.c$ is persisted after $N.d$ and $N.e$, \sys splits this part of $M$'s instance subgraph into Epoch 2 and Epoch 3.

The graph contains the instance subgraphs of $M$ (represented by the colored nodes for updates to $M$) and $N$ ($N.d$ and $N.e$).
The three distinct update regions for $M$ (Epochs 1--3) are only all observable when considering nodes outside of $M$'s instance subgraph. In particular, the boundary between Epochs 2 and 3 can only be observed when considering the update to $N.d$ and $N.e$ from $N$'s instance subgraph.
Consequently, \sys breaks each instance subgraph temporally into \textit{epoch}s by using information from the full persistence graph (\cref{subsec:step-1}) to place program updates that occur close-together in time into the same \textit{epoch subgraph}.
}

\section{Implementation}
\label{sec:implementation}
\label{sec:impl}

We implement \sys in $\sim$12,000 lines of C++ (counted by SLOCCount~\cite{wheeler2001sloccount}).
For \posixBased applications, \sys uses a tracing tool built using Intel Pin~\cite{pin}, a dynamic instrumentation library, that traces all the system calls performed by the application and the memory stores to memory-mapped files.
For efficiency, the syscall tracer only traces syscalls that persist data to disk (e.g. write, rename, unlink) and semantic syscalls that ensure correctness (e.g. open and close).
We compile all the \posixBased applications we tested under debug mode so that the tracer can also output a backtrace for each operation in the execution.

For \mmioBased applications, \sys{} uses pmemcheck~\cite{pmemcheck}, a valgrind~\cite{nethercote2007valgrind} tool for tracing memory operations.
\sys performs program analysis that derives type and instance information based on LLVM-13~\cite{lattner2004llvm,llvm13}.
All \mmioBased applications are also compiled in debug mode to retain the necessary debugging information.

Each crash-consistency bug reported in \sys contains debugging information for the user, including the specific buggy ordering of operations, the corresponding backtraces, and output from the checker program. 
The developers could use the backtrace information to find the source code location of the crash-consistency bugs. 
They can also leverage the buggy ordering of operations to reason about the correct update ordering. 
\sys also provides visualization of subgraphs with annotations for operations to aid developers during debugging.

\section{Evaluation}
\label{sec:evaluation}
\label{sec:eval}

We now evaluate the effectiveness and efficiency of \sys.
We aim to answer 3 main research questions in the evaluation section:

\begin{enumerate}[label=\textbullet,leftmargin=*]
    \item \textbf{RQ1:} Could \sys find crash-consistency bugs in production-ready systems? What are the root causes of crash-consistency bugs found by \sys? (\Cref{subsec:eval-overview}) 
    \item  \textbf{RQ2:} How is \sys's scalability compared to existing state-of-the-art crash-consistency testing tools for \mmioBased and \posixBased applications? (\Cref{subsec:scalability-mmio} \& \Cref{subsec:scalability-posix})
    \item \textbf{RQ3:} How effective are \sys's two implementations of \textit{update behaviors-based} heuristic in approximating the ground-truth set of correlated crash states? (\Cref{subsec:justification}) 
\end{enumerate}

In addition, we provide code coverage comparison and case studies for new crash-consistency bugs found in \mmioBased and \posixBased applications respectively (\Cref{subsec:eval-code-coverage}, \Cref{subsec:eval-hse-case-study} \& \Cref{subsec:eval-rocksdb-case-study}).

\paragraph{Evaluation targets} 
 We evaluate \sys with a variety of libraries and real-world applications.
 For \posixBased applications, we evaluate on \numPOSIXApps production-ready applications: LevelDB~\cite{leveldb}, RocksDB~\cite{rocksdb} and WiredTiger~\cite{wiredtiger}. 
 LevelDB and RocksDB are two widely-used high-performance databases, while WiredTiger serves as the default storage engine for the MongoDB database~\cite{mongodb}.

For \mmioBased applications, we test \numMMIOServerApps  production-ready applications: a PM port of memcached~\cite{PMMemcached} and Redis~\cite{PMRedis} (two popular memory caching services~\cite{carlson2013redis, soliman2013getting}); HSE (Heterogeneous-Memory Storage Engine)~\cite{hse}, a key-value storage engine; and LevelDB~\cite{leveldb} and RocksDB~\cite{rocksdb} with mmap-write enabled.
To compare with other existing crash-consistency testing tools on \mmioBased applications, we also choose \numMMIOMicroApps MMIO microbenchmarks.
We test \numPmdkMicroApps example persistent data structures provided by PMDK~\cite{pmdk}, \numWitcherApps persistent key-value indices~\cite{nam2019write, hwang2018endurable, zuo2018level, lee2017wort}, as well as \numRecipeApps persistent data structures from RECIPE~\cite{recipeRepo,lee2019recipe,witcher-repo}.

\paragraph{Evaluation setup} 
We evaluate our test targets using their provided unit tests and stress tests, with minor modifications to adapt to \sys{}'s workflow. 
For \posixBased applications, we run our experiments on a server with an Intel Xeon Platinum 8480+ CPU (2.00 GHz) and 1024GB of DDR5 DRAM (2200 MHz).
For \mmioBased applications, we run our experiments on a server with an Intel Xeon Gold 6230 CPU (2.10 GHz), 4$\times$128GB Intel Optane Series 100 Pmem DIMMs, and 256GB of DDR4 DRAM (2667 MHz).

\begin{table}[t]
    \centering
    \footnotesize

    \caption{\sys's main testing results. \sys finds \totalServerCCBugs bugs (\newServerCCBugs new) in \numServerApps production-ready systems.}
    
    \hfill

\begin{tabular}{|c|c|cccc|c|}
\hline
\multirow{2}{*}{\textbf{App. Type}} & \multirow{3}{*}{\textbf{App. Name}} & \multicolumn{4}{c|}{\textbf{Root Causes}} & \multirow{2}{*}{\begin{tabular}[c]{@{}c@{}} \textbf{Total} \textbf{Bugs} \\ (\textbf{New})\end{tabular}} \\ 
\cline{3-6} 
& & \multicolumn{1}{c|}{\rotatebox{0}{Atomicity}} & \multicolumn{1}{c|}{\rotatebox{0}{Ordering}} & \multicolumn{1}{c|}{\rotatebox{0}{\makecell{Unpersisted Data}}} & \multicolumn{1}{c|}{\rotatebox{0}{\makecell{Failure Recovery}}} & \\ 
\hline 
\multirow{5}{*}{\textbf{\mmioBased}} & HSE & \multicolumn{1}{c|}{} & \multicolumn{1}{c|}{} & \multicolumn{1}{c|}{} & \multicolumn{1}{c|}{2 (1)} & 2 (1) \\ 
\cline{2-7} 
& Memcached & \multicolumn{1}{c|}{} & \multicolumn{1}{c|}{2 (1)} & \multicolumn{1}{c|}{1 (1)} & & 3 (2) \\ 
\cline{2-7} 
& Redis & \multicolumn{1}{c|}{1 (1)} & \multicolumn{1}{c|}{} & \multicolumn{1}{c|}{} & & 1 (1) \\ 
\cline{2-7} 
& LevelDB-MMIO                                                    & \multicolumn{1}{c|}{}          & \multicolumn{1}{c|}{3}         & \multicolumn{1}{c|}{}                                                    &                                                            & 3 \\
\cline{2-7} 
& RocksDB-MMIO                                                    & \multicolumn{1}{c|}{}          & \multicolumn{1}{c|}{1 (1)}         & \multicolumn{1}{c|}{}                                                    &                                                            & 1 (1)                                                                           \\ \hline \hline
\multirow{3}{*}{\textbf{\posixBased}} 
& LevelDB-POSIX                                                    & \multicolumn{1}{c|}{}          & \multicolumn{1}{c|}{1}         & \multicolumn{1}{c|}{2}                                                    &                                                            & 3 \\
\cline{2-7} 
& RocksDB-POSIX & \multicolumn{1}{c|}{} & \multicolumn{1}{c|}{1 (1)} & \multicolumn{1}{c|}{\add{2}} & & \add{3 (1)} \\ 
\cline{2-7} 
& WiredTiger & \multicolumn{1}{c|}{1 (1)} & \multicolumn{1}{c|}{1} & \multicolumn{1}{c|}{} & & 2 (1) \\ 
\hline 
\hline 
\textbf{Total} & & \multicolumn{1}{c|}{\textbf{2 (2)}} & \multicolumn{1}{c|}{\textbf{9 (3)}} & \multicolumn{1}{c|}{\add{\textbf{5 (1)}}} & \multicolumn{1}{c|}{\textbf{2 (1)}} & \textbf{\textbf{18 (7)}} \\ 
\hline
\end{tabular}

    \label{table:bugs_found}

    \vspace{-1em}
\end{table}

\subsection{RQ1: Bugs Detected by \sys{} and Root Causes}
\label{subsec:eval-overview}

We present the results of \sys's testing in \cref{table:bugs_found}. 
\sys found \totalServerCCBugs bugs across the \numServerApps production-ready systems. 
In \mmioBased applications, \sys finds \totalMMIOServerCCBugs total bugs (\newMMIOServerCCBugs new).
In \posixBased applications, \sys finds \totalPOSIXCCBugs total bugs (\newPOSIXCCBugs new).
We reported the \newServerCCBugs new bugs to their project maintainers, of which \confirmedServerBugs have been confirmed at the time of this writing (all except for the \newMCDBugs memcached, \newRedisBugs Redis, 2 RocksDB bugs). 

We have reported the new RocksDB bugs to the developers with detailed information including the program trace and testing results from crash states that are crash-inconsistent. 
For the two new memcached bugs and the one new Redis bug, we have reported to the developers and also opened pull requests for bug fixes after identifying the root cause of the bugs.

We group the root causes of the crash-consistency bugs found by \sys into four categories: \emph{Atomicity}, \emph{Ordering}, \emph{Unpersisted Data}, and \emph{Failure Recovery}.
An \emph{Atomicity} bug occurs when an update behavior is \emph{assumed} to complete atomically, but improperly leaves a partially-updated state upon a crash.
An \emph{Ordering} bug occurs when an application \emph{assumes} that an update behavior will become persistent in a particular order when undesirable orderings are actually allowed by the underlying storage system.
An \emph{Unpersisted Data} bug occurs when an application \emph{assumes} all data will be made persistent, but a failure to issue explicit ordering calls leads to lost updates.
Finally, a \emph{Failure Recovery} bug is when an application successfully recognizes an inconsistency, but an error in the crash recovery procedure makes the applications unrecoverable.
Among all of the crash-consistency bugs found by \sys, \emph{Ordering} is the most frequent root cause (9/18).

This suggests that developers should highly scrutinize the design of atomic update protocols.
For example, while \stt{valid} flags are commonly used to indicate that a field is not ready to be read, persistent applications often fail to properly ensure that, in \emph{all} cases, the update to \stt{valid} is issued only after all data updates are persisted, and \stt{valid} is subsequently persisted before any future modifications to the data structure.

\subsection{RQ2-1: Scalability of Representative Testing in \mmioBased Applications}
\label{subsec:scalability-mmio}

We evaluate the scalability of representative testing by comparing how quickly it finds bugs compared to state-of-the-art crash-consistency testing tools.

\add{
With respect to \mmioBased applications, we compare against Jaaru\footnote{
\add{Jaaru will stop if it detects a crash-consistency bug during execution, and we extend it to resume the execution when such a case is encountered.
We believe that our comparison between \sys and Jaaru is fair, as the strategy of running a given workload until all bugs have been found is a common approach widely adopted by prior works~\cite{ALICE, fu2022durinn, fu2021witcher}.}
}~\cite{gorjiara2021jaaru}, the state-of-the-art MMIO model checker that employs DPOR crash-state reduction techniques.
Jaaru's original artifact skips library and data structure initialization---however, we find some early crash-consistency bugs that occur during initialization.
Therefore, we evaluate Jaaru both including (``Jaaru'') and excluding (``Jaaru-NoInit'') the initialization code.
Since Jaaru has compatibility issues with newer versions of PMDK, we also implemented Jaaru's DPOR algorithm (``DPOR'') in \sys{}'s model checker to more clearly compare it with representative testing (``RepTest'').
We were also unable to run Jaaru on production-ready \mmioBased systems\footnote{
    The \allServerBugs bugs occur early in these application's executions, so both ``RepTest'' and ``DPOR'' configurations of \sys{} find all \allServerBugs bugs within 11 minutes.}, so
we instead compared with Jaaru on \numMMIOMicroApps MMIO microbenchmarks that include persistent data structures and key-value indices.
}

\begin{figure*}[t]
    \begin{subfigure}{0.43\textwidth}
        \centering
        \includegraphics[width=\columnwidth]{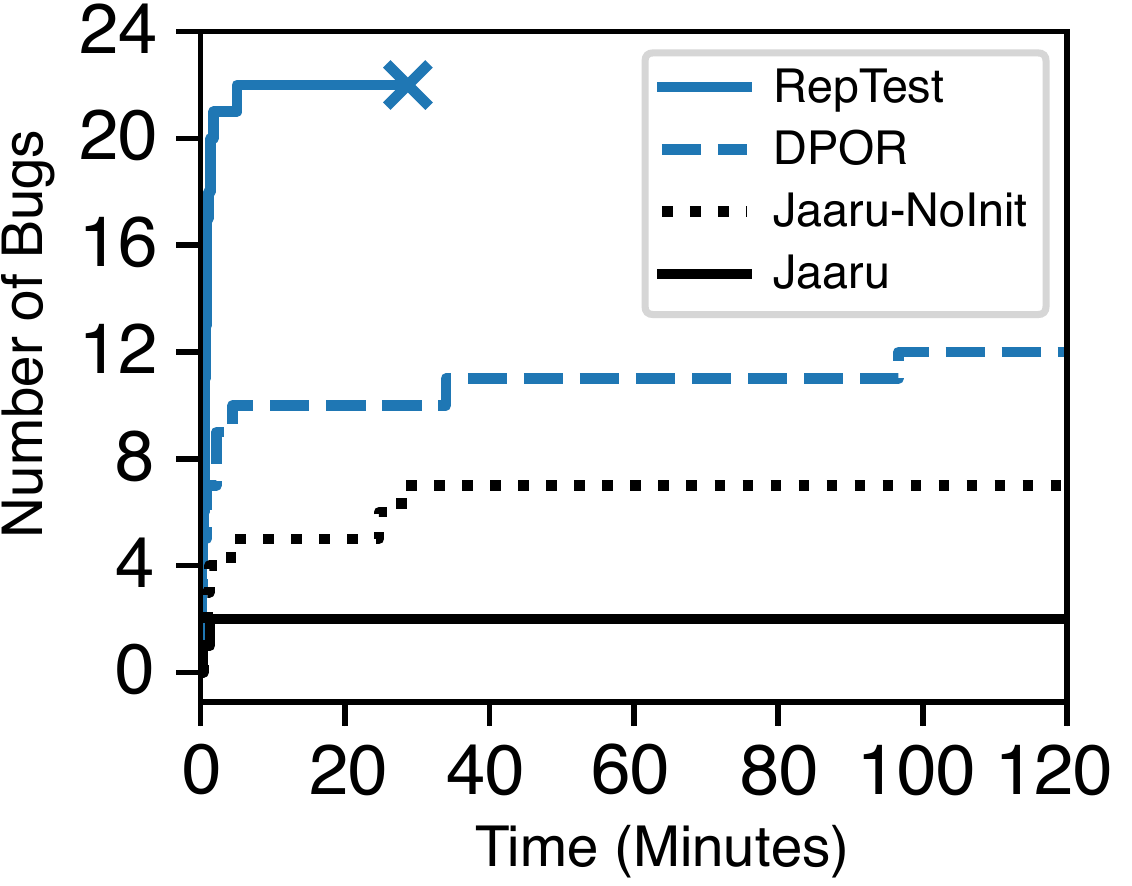}
        \subcaption{PMDK Data Structures}
        \label{subfig:pmdk_bug_graph}
    \end{subfigure}
    \hfill
    \begin{subfigure}{0.43\textwidth}
        \centering
        \includegraphics[width=\columnwidth]{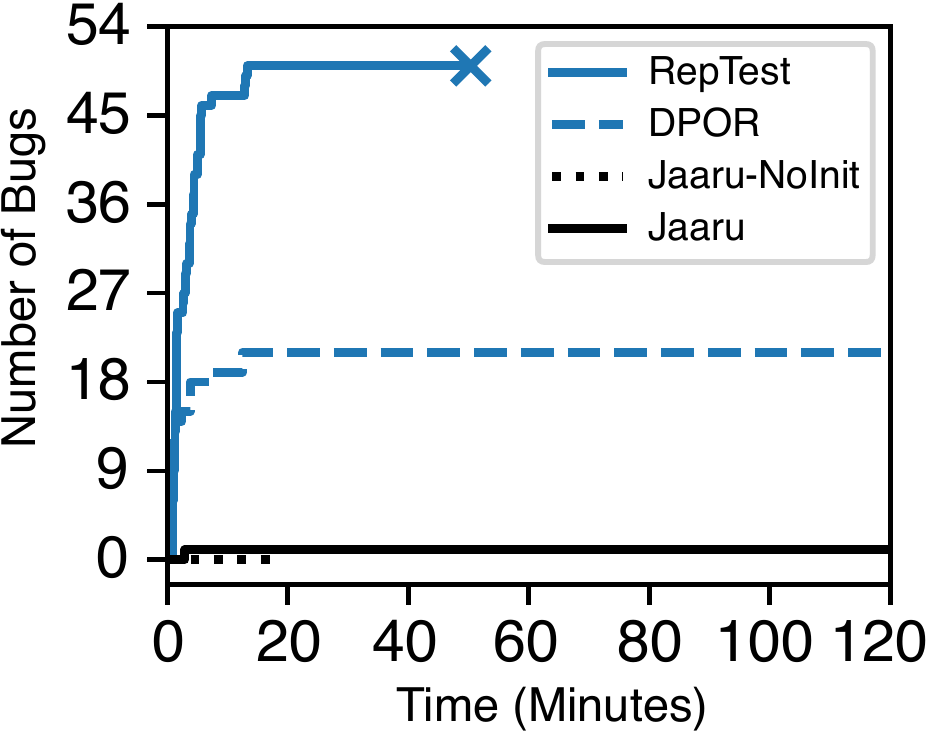}
        \subcaption{Key-Value Indices}
        \label{subfig:kv_bug_graph}
    \end{subfigure}
    \vfill 
    \begin{subfigure}{0.43\textwidth}
        \centering
        \includegraphics[width=\columnwidth]{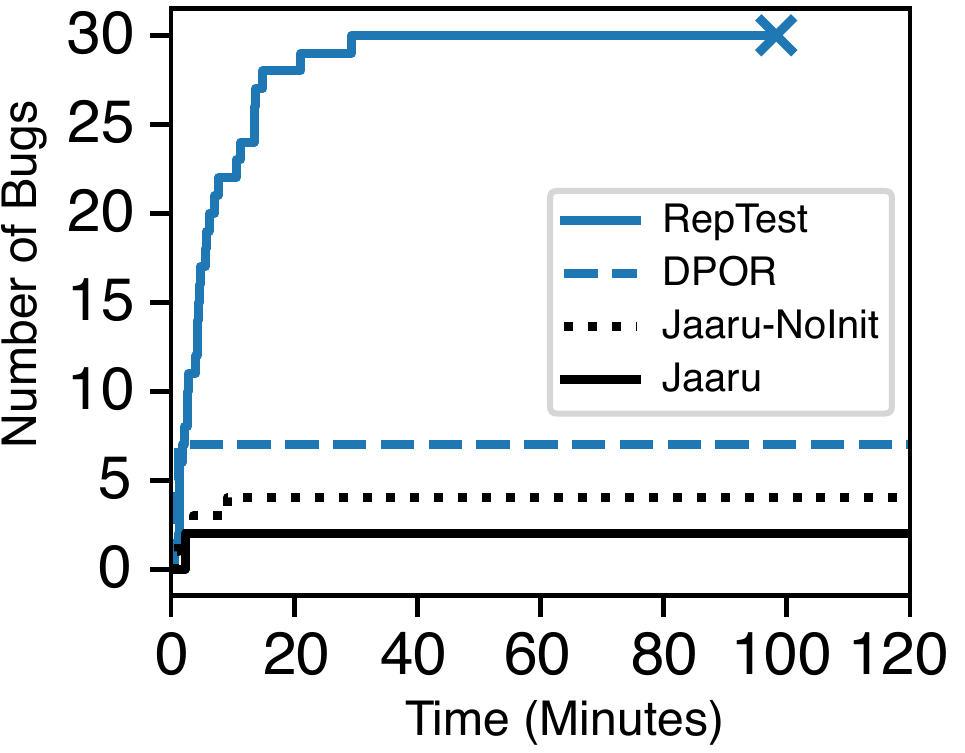}
        \subcaption{RECIPE Indices}
        \label{subfig:recipe_bug_graph}
    \end{subfigure}
    \hfill
    \begin{subfigure}{0.43\textwidth}
        \centering
        \includegraphics[width=\columnwidth]{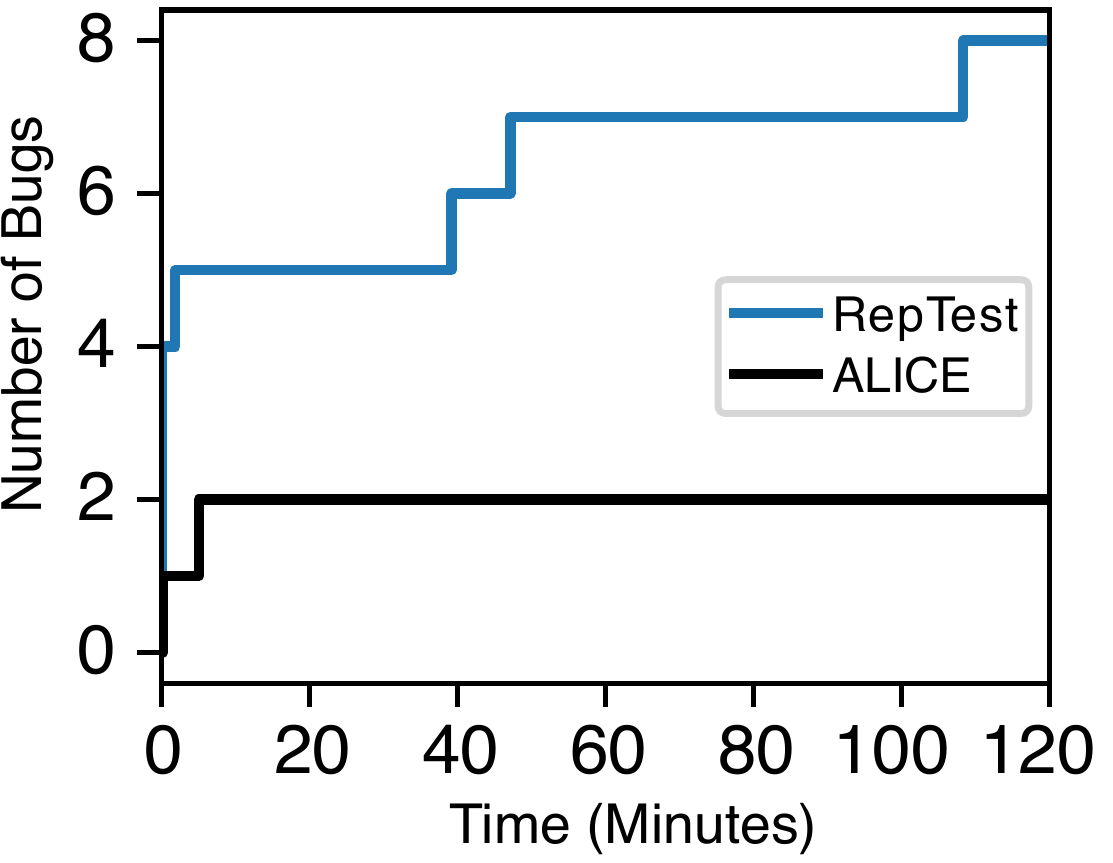}
        \subcaption{\posixBased Applications }
        \label{subfig:posix_bug_graph}
    \end{subfigure}
    
    \caption{\add{A comparison of  \sys{} (``RepTest'') to  baselines on number of crash-consistency bugs found over testing time. \Cref{subfig:pmdk_bug_graph} to \cref{subfig:recipe_bug_graph} are for \mmioBased microbenchmarks and
    \Cref{subfig:posix_bug_graph} is for \posixBased production-ready applications introduced in \cref{table:bugs_found}.
     The $\times$ marker indicates the completion of testing.}}
    \label{fig:results-mmio}
\end{figure*}

\Cref{subfig:pmdk_bug_graph} to \cref{subfig:recipe_bug_graph} shows the bugs found over time for three categories of  microbenchmarks in \mmioBased applications.
Each graph sums the results of the individual tests. 
Representative testing outperforms existing approaches in all benchmarks.
\sys{} finds all of the reported bugs before the time limit (28 minutes for PMDK data structures, 50 minutes for Key-Value indices, 98 minutes for RECIPE indices).
In contrast, none of the baselines finds all of the bugs within the time limit.
The DPOR baseline outperforms both Jaaru configurations but finds roughly half of the bugs found by representative testing.
Jaaru performs even worse, as the Jaaru-NoInit and Jaaru baselines find ~$12\%$ and ~$5\%$ of the bugs found by representative testing, respectively.
\add{
Given an infinite testing budget, the baselines should find all bugs found by \sys{}; however, \sys{} scales crash-consistency testing to large applications and finds crash-consistency bugs quickly.
}

\Cref{table:bugs_found_micro} shows the number of crash-consistency bugs found in each MMIO microbenchmark, categorized by root causes.
\sys found \totalMMIOMicroCCBugs bugs across the \numMMIOMicroApps microbenchmarks.
In our testing, we find \newMMIOMicroCCBugs new bugs: \newPMDKBugs in PMDK's \textit{Array} data structure, \newKVBugs in the persistent key-value indices, \newRecipeBugs in RECIPE data structures.
All of these systems were previously tested by prior work~\cite{fu2021witcher,fu2021witcher,liu2021pmfuzz,neal2020agamotto,liu2020cross}, yet these new bugs were not found by any prior tool.
We reported the \newMMIOMicroCCBugs new bugs to their project maintainers, all of which have been confirmed by the developers.

We further compare the number of crash-consistency bugs found by \sys with Witcher~\cite{fu2021witcher}, as it has previously made public all the crash-consistency bugs reported.
Pathfinder is able to find all other than two crash-consistency bugs reported in Witcher: Bug \#28~\cite{bug28witcher} in BwTree and Bug \#36~\cite{bug36witcher} in P-HOT. 
Both of these bugs cause the crash-recovery code to leak memory instead of returning an error status. 
Pathfinder tests erroneous operations but does not identify them as failures. 
However, Pathfinder finds 52 bugs that Witcher fails to report.
\add{
Witcher is a pattern-based detector as it prunes the crash-state space by only testing states that are related to “guardian protection”, a programming pattern the authors claimed to be common in MMIO programming. 
This leads to false negatives when testing programs that do not follow this programming behavior.
}

\begin{table}[t]
    \centering
    \footnotesize
    \caption{\sys's testing results on MMIO microbenchmarks.}
    
    \hfill

\begin{tabular}{|c|c|cccc|c|}
\hline
\multirow{2}{*}{\textbf{App. Type}} & \multirow{3}{*}{\textbf{App. Name}} & \multicolumn{4}{c|}{\textbf{Root Causes}} & \multirow{2}{*}{\begin{tabular}[c]{@{}c@{}} \textbf{Total} \textbf{Bugs} \\ (\textbf{New})\end{tabular}} \\ 
\cline{3-6} 
& & \multicolumn{1}{c|}{\rotatebox{0}{Atomicity}} & \multicolumn{1}{c|}{\rotatebox{0}{Ordering}} & \multicolumn{1}{c|}{\rotatebox{0}{\makecell{Unpersisted Data}}} & \multicolumn{1}{c|}{\rotatebox{0}{\makecell{Failure Recovery}}} & \\ 
\hline 
\multirow{6}{*}{\begin{tabular}[c]{@{}c@{}}\textbf{PMDK Data} \\ \textbf{Structures}\end{tabular}} 
& Array & \multicolumn{1}{c|}{2 (1)} & \multicolumn{1}{c|}{5 (5)} & \multicolumn{1}{c|}{} & & 7 (6) \\ 
\cline{2-7} 
& BTree & \multicolumn{1}{c|}{4} & \multicolumn{1}{c|}{} & \multicolumn{1}{c|}{} & & 4 \\ 
\cline{2-7} 
& CTree & \multicolumn{1}{c|}{1} & \multicolumn{1}{c|}{} & \multicolumn{1}{c|}{} & & 1 \\ 
\cline{2-7} 
& \begin{tabular}[c]{@{}c@{}}Hashmap (Atomic)\end{tabular} & \multicolumn{1}{c|}{4} & \multicolumn{1}{c|}{} & \multicolumn{1}{c|}{} & & 4 \\ 
\cline{2-7} 
& \begin{tabular}[c]{@{}c@{}}Hashmap (TX)\end{tabular} & \multicolumn{1}{c|}{} & \multicolumn{1}{c|}{} & \multicolumn{1}{c|}{1} & \multicolumn{1}{c|}{1} & 2 \\ 
\cline{2-7} 
& RBTree & \multicolumn{1}{c|}{1} & \multicolumn{1}{c|}{} & \multicolumn{1}{c|}{3} & & 4 \\ 
\hline \hline
\multirow{5}{*}{\begin{tabular}[c]{@{}c@{}}\textbf{Key-Value} \\ \textbf{Indices}\end{tabular} } 
& CCEH & \multicolumn{1}{c|}{4 (2)} & \multicolumn{1}{c|}{} & \multicolumn{1}{c|}{} & & 4 (2) \\ 
\cline{2-7} 
& Fast Fair & \multicolumn{1}{c|}{7 (3)} & \multicolumn{1}{c|}{} & \multicolumn{1}{c|}{} & & 7 (3) \\ 
\cline{2-7} 
& Level Hash & \multicolumn{1}{c|}{13 (6)} & \multicolumn{1}{c|}{15 (5)} & \multicolumn{1}{c|}{6 (6)} & & 34 (17) \\ 
\cline{2-7} 
& WOART & \multicolumn{1}{c|}{3 (2)} & \multicolumn{1}{c|}{} & \multicolumn{1}{c|}{} & & 3 (2) \\ 
\cline{2-7} 
& WORT & \multicolumn{1}{c|}{2 (2)} & \multicolumn{1}{c|}{} & \multicolumn{1}{c|}{} & & 2 (2) \\ 
\hline \hline
\multirow{5}{*}{\begin{tabular}[c]{@{}c@{}}\textbf{RECIPE} \\ \textbf{Indices}\end{tabular} } 
& P-ART & \multicolumn{1}{c|}{4 (2)} & \multicolumn{1}{c|}{} & \multicolumn{1}{c|}{} & & 4 (2) \\ 
\cline{2-7} 
& P-BwTree & \multicolumn{1}{c|}{3 (2)} & \multicolumn{1}{c|}{} & \multicolumn{1}{c|}{} & & 3 (2) \\ 
\cline{2-7} 
& P-CLHT & \multicolumn{1}{c|}{8 (5)} & \multicolumn{1}{c|}{1} & \multicolumn{1}{c|}{3 (1)} & \multicolumn{1}{c|}{1 (1)} & 13 (7) \\ 
\cline{2-7} 
& P-HOT & \multicolumn{1}{c|}{5 (5)} & \multicolumn{1}{c|}{2} & \multicolumn{1}{c|}{} & & 7 (5) \\ 
\cline{2-7} 
& P-Masstree & \multicolumn{1}{c|}{3 (1)} & \multicolumn{1}{c|}{} & \multicolumn{1}{c|}{} & & 3 (1) \\ 
\hline
\hline
\textbf{Total} & & \multicolumn{1}{c|}{\textbf{64 (31)}} & \multicolumn{1}{c|}{\textbf{23 (10)}} & \multicolumn{1}{c|}{\textbf{13 (7)}} & \multicolumn{1}{c|}{\textbf{2 (1)}} & \textbf{102 (49)} \\ 
\hline
\end{tabular}

    \label{table:bugs_found_micro}
\end{table}

\subsection{RQ2-2: Scalability of Representative Testing in \posixBased Applications}
\label{subsec:scalability-posix}

For \posixBased applications, we compare against ALICE~\cite{ALICE}, a crash-consistency testing tool commonly used by the database community to systematically explore crash states in the program execution.
We enable multi-threading in ALICE to improve its performance during trace parsing and model checking.
We use \sys{} and the baseline testing strategies on each of our testing targets and count the number of bugs found by each approach over time. 
We set a maximum time limit of 2 hours for all application categories.

\Cref{subfig:posix_bug_graph} shows the crash-consistency bugs found over time for \posixBased production-ready systems (LevelDB, RocksDB and WiredTiger).
\sys finds all \totalPOSIXCCBugs bugs within the \POSIXTime time limit.
In contrast, ALICE finds only 2 bugs within the time limit.
The 2 bugs found by ALICE are from LevelDB, which uses a small workload of $\sim$10 key-value pairs insertions.
However, our other \posixBased application workloads are larger stress tests that perform between 1,000--100,000 operations, which ALICE fails to scale to.

\subsection{RQ3: \emph{Update Behaviors-Based} Heuristic Efficacy}
\label{subsec:justification}

\begin{table}[t]
    \centering
    \footnotesize
    \caption{A comparison of \sys{} to baselines for number of correlated crash states and total crash states.
    }
    
    \begin{tabular}{|c|c|ccc|ccc|}
\hline
\multirow{2}{*}{\textbf{App. Type}} & \multirow{2}{*}{\textbf{App. Name}} & \multicolumn{3}{c|}{\textbf{Correlated Crash States}} & \multicolumn{3}{c|}{\textbf{Total Crash States}} \\ 
\cline{3-8} 
& & \multicolumn{1}{c|}{\sys{}} & \multicolumn{1}{c|}{Persevere} & \multicolumn{1}{c|}{Exhaustive} & \multicolumn{1}{c|}{\sys{}} & \multicolumn{1}{c|}{Persevere} & \multicolumn{1}{c|}{Exhaustive} \\ 
\hline 
\multirow{3}{*}{\textbf{MMIO}} 
& HSE & \multicolumn{1}{c|}{4} & \multicolumn{1}{c|}{/} & \multicolumn{1}{c|}{N/A\footnotemark} & \multicolumn{1}{c|}{88} & \multicolumn{1}{c|}{/} & \multicolumn{1}{c|}{4,414} \\ 
\cline{2-8} 
& Memcached & \multicolumn{1}{c|}{231} & \multicolumn{1}{c|}{/} & \multicolumn{1}{c|}{7,743} & \multicolumn{1}{c|}{432} & \multicolumn{1}{c|}{/} & \multicolumn{1}{c|}{9,161} \\ 
\cline{2-8} 
& Redis & \multicolumn{1}{c|}{661} & \multicolumn{1}{c|}{/} & \multicolumn{1}{c|}{1,076} & \multicolumn{1}{c|}{977} & \multicolumn{1}{c|}{/} & \multicolumn{1}{c|}{10,004} \\ 
\cline{2-8} 
\hline \hline
\multirow{3}{*}{\textbf{POSIX}} 
& LevelDB & \multicolumn{1}{c|}{99} & \multicolumn{1}{c|}{128} & \multicolumn{1}{c|}{478} & \multicolumn{1}{c|}{3,597} & \multicolumn{1}{c|}{1,936,443} & \multicolumn{1}{c|}{2,093,865} \\ 
\cline{2-8} 
& RocksDB & \multicolumn{1}{c|}{1,120} & \multicolumn{1}{c|}{20,070} & \multicolumn{1}{c|}{172,331} & \multicolumn{1}{c|}{37,478} & \multicolumn{1}{c|}{10,701,527} & \multicolumn{1}{c|}{16,477,850} \\ 
\cline{2-8} 
& WiredTiger & \multicolumn{1}{c|}{266} & \multicolumn{1}{c|}{337} & \multicolumn{1}{c|}{1,315} & \multicolumn{1}{c|}{1,252} & \multicolumn{1}{c|}{2,084} & \multicolumn{1}{c|}{23,058} \\ 
\hline 
\end{tabular}

    \label{table:crash_state}
\end{table}
\footnotetext{The exhaustive model checker fails to detect HSE bugs within the 2-hours limit.}

To show the efficacy of our update behaviors-based heuristics, we compare the number of correlated crash states and total crash states tested by \sys{} with baselines.
We define a crash state to be a correlated crash state associated with a crash-consistency bug, if the crash schedule that generates this crash state covers the source code location where the root cause of the crash-consistency bug is at.
We select two baselines: Persevere\footnote{Persevere is only able to test simple, self-contained C syscall program and does not support testing production-ready systems, which is a drawback confirmed by the developers. Thus, we does not report efficiency comparison with Persevere in \Cref{subsec:scalability-posix}. Instead, we re-implement Persevere's algorithm and report crash state comparison.}~\cite{persevere}, the state-of-the-art POSIX data persistency model checker based on DPOR techniques, and an exhaustive model checker that enumerates all possible crash states generated by a program trace.

\Cref{table:crash_state} shows the number of correlated crash states and total crash states tested for \sys{} and baselines.
We sum up crash states from all the crash-consistency bugs in each target application.
For test efficiency, we set the time limit of testing both \posixBased and \mmioBased applications to be 2 hours.
The exhaustive model checker presents a lower bound on the number of ground-truth crash states that will be tested.
In all applications, \sys{} is able to detect crash-consistency bugs with the minimal number of correlated crash states and total crash states.
Compared with the exhaustive baseline, \sys{} is able to reduce the number of correlated crash states by 38\% to 99\% and reduce the number of total crash states by 90\% to 99\%.
Compared with Persevere~\cite{persevere}, \sys{} is able to reduce the number of correlated crash states by 21\% to 94\% and reduce the number of total crash states by 40\% to 99\%.
This demonstrates the effectiveness of our two \emph{update behaviors-based} heuristic implementations at pruning the crash-state space.

\add{
\subsection{Code Coverage Comparison}
\label{subsec:eval-code-coverage}

\begin{figure*}[t]
    \begin{subfigure}{0.32\textwidth}
        \centering
        \includegraphics[width=\columnwidth]{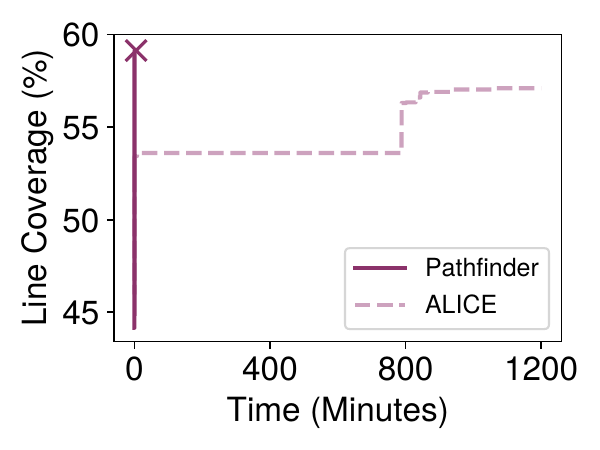}
        \subcaption{LevelDB}
        \label{subfig:code-coverage-leveldb}
    \end{subfigure}
    \begin{subfigure}{0.32\textwidth}
        \centering
        \includegraphics[width=\columnwidth]{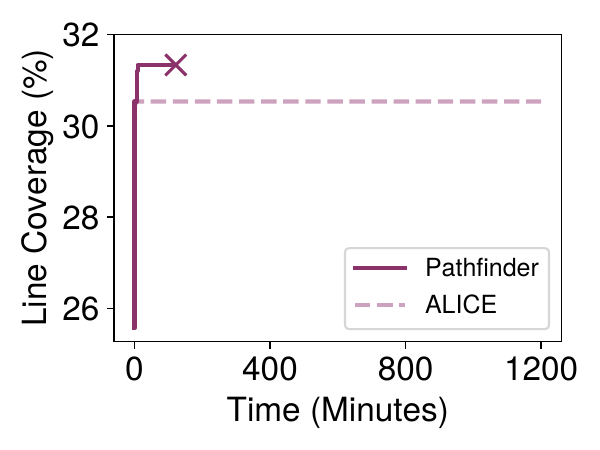}
        \subcaption{RocksDB}
        \label{subfig:code-coverage-rocksdb}
    \end{subfigure}
    \begin{subfigure}{0.32\textwidth}
        \centering
        \includegraphics[width=\columnwidth]{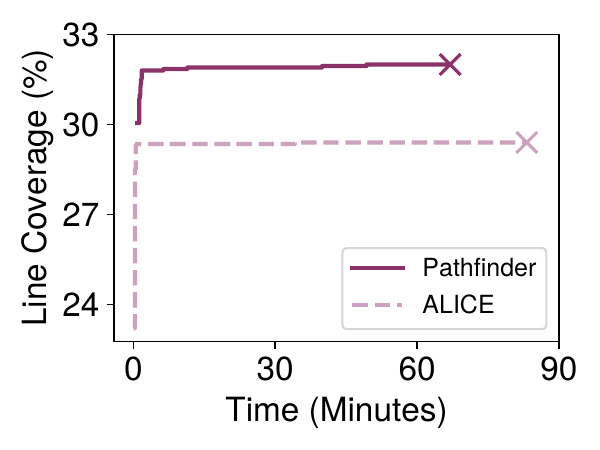}
        \subcaption{WiredTiger}
        \label{subfig:code-coverage-wiredtiger}
    \end{subfigure}
    
    \caption{Code coverage comparison of \sys{} and ALICE~\cite{ALICE}.}
    \label{fig:code-coverage}
    \vspace{-1em}
\end{figure*}

\Cref{fig:code-coverage} presents a comparison of code coverage between \sys{} and ALICE~\cite{ALICE} across all tested \posixBased production-ready systems, measured in terms of line coverage. For each system, we execute all workloads associated with the crash-consistency bugs identified by \sys{} and compute the average line coverage at each time step. The maximum execution time is capped at 20 hours (1200 minutes). The baseline coverage represents the line coverage after executing the workloads but before running crash-recovery procedures. Across all tested systems, \sys{} consistently achieves higher code coverage than ALICE and does so at a significantly faster speed.
}

\subsection{MMIO Case Study: Incorrect crash recovery in HSE}
\label{subsec:eval-hse-case-study}

\Cref{lst:hse_new_bug} illustrates a new crash-consistency bug found in HSE~\cite{hse} by \sys.
The storage engine defines a custom log header data structure \stt{mdc\_loghdr} with 4 data fields for its data file \stt{mdc\_file}.
During log header updates in function \stt{mdc\_loghdr\_pack}, an untimely crash may result in the update to checksum \stt{crc} field not being persisted (line~\ref{line:crash_site_hse}).
Then during the opening of the data file in \stt{mdc\_file\_open}, the validation procedure in \stt{mdc\_file\_validate} detects that a crash has happened and returns an error.
The correct crash recovery process in this case is to keep the pointer for data file \stt{mfp} so that the storage engine could erase the log header and try reopening again.
However, the developers incorrectly free the file pointer in \stt{err\_exit1}, causing the file open to fail.

\begin{figure}[t]
    \centering
    \begin{minipage}{.48\textwidth}
        \centering                \begin{lstlisting}[language=C,style=customnew,caption={Code snippet of a new crash-consistency bug in HSE~\cite{hse} found by \sys. The log header may be partially updated during a crash. During crash recovery the log handle may be erroneously freed, causing the database open to fail. },label={lst:hse_new_bug},escapechar=|]
// MDC Log header
struct mdc_loghdr {
  uint32_t vers;
  uint32_t magic;
  uint32_t gen;
  uint64_t crc; // checksum
} 
merr_t omf_mdc_loghdr_pack(
  struct mdc_loghdr *lh, ...) {
  omf_set_lh_vers(lhomf, lh->vers);
  omf_set_lh_magic(lhomf, lh->magic);
  omf_set_lh_gen(lhomf, lh->gen);
  |\hl{/* Crash occurs here! */} \label{line:crash_site_hse}|
  ...
  omf_set_lh_crc(lhomf, crc);
}
merr_t mdc_file_open(...) {
  struct mdc_file *mfp;
  ...
  err = mdc_file_validate(mfp, ...);
  if (err) { goto err_exit1; }
  ...
err_exit1:
  free(mfp);
}
\end{lstlisting}
    \end{minipage}%
    \begin{minipage}{.48\textwidth}
    \centering
\begin{lstlisting}[language=C,style=customnew,caption={Code snippet of a new crash-consistency bug in RocksDB~\cite{rocksdb} found by \sys. The write operations during MemTable flush may be reordered if they happen to be on the block boundary, causing RocksDB to have inconsistent behavior when printing out key and values. },label={lst:rocksdb_new_bug},escapechar=|]
Status DBImpl::Write(...) {
  // Register with write thread
  Status status = write_thread_.
    JoinBatchGroup(&w);
  // Write to Write-Ahead Log
  status = WriteToWAL(
    write_options, w.batch, &log_used);
  // Write to MemTable
  status = WriteBatch::InsertInto(...);
  ...
}
Status DBImpl::FlushMemTable(...) {
  // Ensure we have a MemTable to flush
  MemTable* mem = cfd->imm()->current();
  // Build the Table (SSTable)
  TableBuilder* builder;
  // Iterator to read from MemTable
  Iterator* mem_iter = mem->NewIter();
  for (mem_iter->SeekToFirst();...) {  |\label{line:flush_rocksdb}|
    Slice key = mem_iter->key();
    Slice value = mem_iter->value();
    builder->Add(key, value);
    |\hl{/* Crash occurs here! */} \label{line:crash_site_rocksdb}|
  }
}
\end{lstlisting}
    \end{minipage}
\vspace{-2em}
\end{figure}

We observe that this crash-consistency bug could be detected by generating crash states that are associated with this single data structure \stt{mdc\_loghdr}.
Prior crash-consistency testing tools for \mmioBased applications are unable to find this bug as they cannot scale to large systems like HSE.

\subsection{POSIX Case Study: Inconsistent behavior in RocksDB}
\label{subsec:eval-rocksdb-case-study}
\Cref{lst:rocksdb_new_bug} demonstrates a new crash-consistency bug found in RocksDB~\cite{rocksdb} by \sys.
When the user inserts key-value pairs into the database, function \stt{DBImpl::Write} will be invoked.
The changes will first be written to a Write-Ahead Log (WAL), along with an in-memory MemTable.
When the MemTable is full or a manual flush is triggered, \stt{DBImpl::FlushMemTable} will be called to flush all contents in the MemTable to an immutable SSTable file on disk.
This process will be done iteratively by enumrating contents in the MemTable (line~\ref{line:flush_rocksdb}).
Particularly, function \stt{builder->Add} will use \stt{write} syscalls to persist changes to the disk, where a crash may happen (line~\ref{line:crash_site_rocksdb}).
In most cases this will not lead to inconsistency, as \stt{write} syscalls to the same block cannot be reordered.
However, if two consecutive \stt{write} syscalls happen to be on the block boundary, it is possible that they can be reordered according to ext4 file system model~\cite{Ferrite}.
In this case, RocksDB will still print out newer key-value pairs although older key-value pairs are lost, creating holes in writes.
This is an inconsistent behavior that should not be tolerated according to RocksDB developers~\cite{rocksdb-no-holes}.

We observe that this crash-consistency bug could be detected by generating crash states that are associated with the function \stt{DBImpl::FlushMemTable}.
 A corner case like this can only be stably triggered in a large stress test, while previous crash-consistency testing tools for \posixBased applications fail to scale to such test cases.

\section{Discussion}
\label{sec:discussion}

\paragraph{False negatives} 
As explained in \Cref{subsec:scalability-mmio}, \sys{} has two false negatives from RECIPE data structures~\cite{lee2019recipe}.
A false negative is defined as a known crash-consistency bug reported by prior works in the applications tested that \sys{} fails to find.
Like existing tools, \sys may have false negatives if: (1) the test oracle fails to detect inconsistencies in crash states or (2) an inconsistent crash state is not tested by the groups of representative update behaviors in \sys{}.
(1) can be mitigated by developing an oracle that performs more error-checking, following previous works on crash-consistency validation~\cite{jiang2016crash}.
To mitigate (2), \sys{} prioritizes testing update behaviors from the smallest size to the largest.

\paragraph{False positives}
During our evaluation, we have not observed false positives from \sys{}.
A false positive is defined as a crash-consistency bug reported by \sys{} that does not actually exist in the code.
\sys reports a bug if and only if the crash-recovery program reports an error status. 
Therefore, as long as the crash-recovery program is correctly implemented (as is assumed in this work and prior works~\cite{fu2021witcher, gorjiara2021jaaru}), there should be no false positives in \sys.

\paragraph{Test case generation} \sys can only detect a crash-consistency bug if the operation trace (Step A, \cref{subsec:step-1}) contains an update behavior that evinces the bug. 
\sys currently uses unit tests and stress tests to drive its testing and analysis.
Adopting test generation approaches (e.g., fuzzing~\cite{maier2020aflpp,liu2021pmfuzz,xu2017designing} or symbolic execution~\cite{cadar2008klee,neal2020agamotto}) could increase the number and size of the operation traces for \sys to analyze.
This would increase the number of update behaviors that \sys could test and could potentially find more crash-consistency bugs. 
Such approaches will be complementary but orthogonal to the goal of \sys.

\paragraph{Multi-threaded programs} 
Although \sys{} supports tracing across multiple threads, it cannot systematically detect crash-consistency bugs that arise from concurrency issues due to lack of support for tracing thread synchronization operations.
\sys currently generates crash states by only reordering operations following the persistent happens-before model across threads, as if a mutex exists between context switching. 
This means \sys will generate strictly fewer crash states than is otherwise allowed by concurrent execution.
This assumption may not always hold, but ensures that no false positive is generated from impossible thread interleavings.

\paragraph{Supported applications} 
\sys currently supports testing \posixBased and \mmioBased applications.
For applications that use both syscalls and MMIO, \sys can detect crash-consistency bugs from either syscall-level reordering, or memory operation-level reordering, but not both.
In our extensive survey of prior work and real reported open-source bugs, we have yet to identify crash-consistency bugs that are caused from a combination of syscalls and MMIO.

\paragraph{Synergies with GNNs}
\sys relies on heuristics methods for deriving update behaviors from a full persistence graph (\cref{sec:algorithms}).
These update behavior heuristics are a product of our observations and experience dealing with MMIO and POSIX storage systems, and while they are effective in \sys (\cref{sec:eval}), updates and refinements to the algorithms are limited by human effort. 
To overcome this limitation, representative testing systems could leverage graph neural networks (GNNs) to automatically analyze and find patterns in persistence graphs that can then be used to generate novel update behavior derivation algorithms.
For example, the backtrace and syscall information in \posixBased applications and the data type information in \mmioBased applications can be used to create feature vectors for each node.
Then graphs could be grouped into clusters based on vector distance.
Finally, top-K graphs from each cluster may be selected for testing based on customized sampling methods.
Recent work has used GNNs to perform code similarity analysis and detect bugs (\cite{luo2022compact,liu2023codeformer}), so we believe this is a promising avenue for future work in representative testing.

\paragraph{Generality of representative testing}
We believe that representative testing can be generalized and applied to applications outside of the MMIO-based and POSIX-based storage applications studied in this paper.
The core observation that enables representative testing is that application developers often write code where semantically-related program updates are grouped together as \textit{update behaviors} and that \textit{representative} update behaviors can be tested in lieu of testing all update behaviors while still exposing the same set of bugs (see \Cref{subsec:general-methodology}).
\add{
We believe that our core observation extends to other domains as well.
}
For example, a survey on concurrency bugs in cloud computing applications~\cite{leesatapornwongsa2016taxdc} found that more than 60\% of distributed concurrency bugs were caused by a single message ordering violation in a protocol---such bugs could be found by representative testing by treating message protocols as update behaviors in the distributed system, then testing possible message orderings within representative instances of the message protocol.

We describe the procedure for applying representative testing to another application domain through the example of distributed system testing:
\begin{enumerate}[leftmargin=*]
    \item \textbf{Define the operation graph.} 
    In \sys, we use the \textit{persistence graph} abstraction to define the ordering relationships between persistence updates (\cref{subsec:step-4}).
    A similar graph abstraction needs to be defined for the new domain to be tested.
    For distributed system testing, the nodes of the graph could be defined to be messages sent between servers, with the edges between those nodes being the happens-before relationships between messages as inferred by prior work~\cite{liu2017dcatch}.

    \item \textbf{Define the update behavior heuristics.}
    To break down the whole-program persistence graph into update behaviors (\cref{subsec:step-3}), \sys uses a set of update behavior derivation algorithms (\cref{sec:algorithms}) primarily based on static source code location.
    \add{
    For distributed system testing, the source code location that sends a message may be used as a graph splitting criteria, or the structure of a message may be used as a feature that distinguishes types of update behaviors.
    }
    
    \item \textbf{Create the mechanisms for tracing and testing.}
    A representative testing system such as \sys needs to be able to trace applications (\cref{subsec:step-1}) to create and analyze update behavior graphs and also needs model checking infrastructure to be able to test update behaviors (\cref{subsec:step-5}).
    For distributed system testing, prior work in distributed system tracing~\cite{liu2017dcatch} and model checking~\cite{leesatapornwongsa2014samc} can be leveraged and modified to allow for testing of specific update behaviors.
\end{enumerate}

\section{Related Work}
\label{sec:related}
\paragraph{Application-level crash-consistency testing}
Prior works have explored application-level crash-consistency testing and attempted to solve crash-state space explosion.
Some prune search space by known buggy patterns~\cite{PACE, torturingdb, CrashMonkey, fu2021witcher}, or try eliminating redundancy in the search space~\cite{FiSC, leesatapornwongsa2014samc,gorjiara2021jaaru}.
However, existing methods sacrifice either scalability or accuracy.
Instead, \sys{} uses representative testing to achieve high scalability and accuracy for crash-consistency testing.

\paragraph{File system crash-consistency testing} 
Many prior systems and tools explore crash-consistency testing in file systems~\cite{mohan2018finding,martinez2017crashmonkey,chen2015using,kim2019finding,fryer2012recon,fryer2014checking,hance2020storage,pillai2017application,jiang2016crash,FiSC,yang2006explode}. 
These approaches primarily leverage either bounded pruning or application-specific customization. 
For example, some prune the search space by bounding the number of system calls they explore~\cite{mohan2018finding,martinez2017crashmonkey}. 
Others only test or verify specific file systems or file system operations~\cite{chen2015using,hance2020storage,keller2014file,yang2006explode,FiSC,kim2019finding,xu2019fuzzing}. 
Finally, some tools only test if applications are using the file system interface correctly~\cite{jiang2016crash}. 
In contrast to these approaches, representative testing only requires a different implementation of the \emph{update behaviors-based} heuristic for different types of applications (\cref{sec:algorithms}).

\paragraph{\mmioBased programming libraries} Many developers of \mmioBased applications use \mmioBased programming libraries~\cite{pmdk,coburn2011nvheaps,volos2011mnemosyne,xu2021clobber,scargall2020libpmemobj,zhang2019pangolin,memaripour2017atomic} rather than re-implementing common durability techniques (e.g., undo-logging). 
While small \mmioBased libraries can be exhaustively tested~\cite{gorjiara2021jaaru}, crash-consistency bugs can arise from API misuse or through specific sequences of API calls which may not be tested in library-only model checking. Since \sys{} effectively prunes the testing space for \mmioBased applications, \sys{} enables a developer to identify crash-consistency bugs in the application code, the library code, and the interface between the two.

\paragraph{Partial-order reduction techniques} 
Partial-order reduction techniques~\cite{godefroid96partial, flanagan2005dynamic} are used by a variety of testing techniques that use stateless model checking~\cite{godefroid1997model} to avoid generating redundant states.
Partial-order reduction accelerates model checking by identifying commutative operation orderings that generate identical states and elides generating such states.
The pruning provided by partial-order reduction alone is insufficient for crash-states spaces, however, as there are still and exponential number of \textit{unique} crash states for exhaustive approaches to test~\cite{gorjiara2021jaaru}.
\sys{} scales beyond such approaches by using representative testing (\cref{sec:rep-testing}).

\section{Conclusion}
\label{sec:conclusion}

The key challenge in crash-consistency testing is that the crash-state space grows exponentially as the number of operations increases. 
Existing methods prune this space by targeting known buggy patterns or skipping identical states, but they compromise coverage and scalability.
Based on our observation of correlated crash states, we developed \sys, a scalable and accurate testing tool that uses an \emph{update behaviors-based} heuristic to approximate representative crash states and effectively prune the crash-state space by testing these representatives.
We used \sys to efficiently discover \totalServerCCBugs bugs (\newServerCCBugs new) in \numServerApps \posixBased and \mmioBased production-ready applications, demonstrating \sys{}'s superior scalability and accuracy over existing methods.

\section*{Acknowledgments}
We would like to thank the anonymous reviewers of OOPSLA for their valuable comments on the paper.
We also thank Shuangrui Ding and Yigong Hu for their insightful feedback.
We thank Hossein Golestani for the contributions to an initial prototype of \sys{}.
This work is generously supported by NSF CAREER Award \#2333885, Intel’s Center for Transformative Server Architectures (TSA), the PRISM Research Center, a JUMP Center cosponsored by SRC and DARPA.

\section*{Data-Availability Statement}

The software that implements techniques described in \Cref{sec:design} to \Cref{sec:algorithms} and supports evaluation results in \Cref{sec:evaluation} is publicly available at \url{https://github.com/efeslab/Pathfinder}.

\bibliographystyle{ACM-Reference-Format}
\bibliography{references}

\appendix

\end{document}